\documentclass[english,floatfix,11pt,superscriptaddress,aps,prd,reprint,showkeys,nofootinbib]{revtex4-2}
\usepackage{amsmath}
\usepackage{amssymb}
\usepackage{amsbsy}
\usepackage{amsfonts}
\usepackage{amsopn}
\usepackage{amstext}
\usepackage{graphicx}
\usepackage[english]{babel}
\usepackage{color}
\usepackage{slashed}
\usepackage{esint}
\usepackage[dvips]{epsfig}
\usepackage[dvips]{graphicx}
\usepackage{float}
\usepackage{units}
\usepackage{textcomp}
\usepackage{wasysym}
\usepackage{hyperref}
\usepackage{slashed}
\usepackage{orcidlink}

\graphicspath{{./}}

\begin{document}

\title{Perturbations and greybody bounds of Euler-Heisenberg black holes surrounded by perfect fluid dark matter}

\author{Fernando M. Belchior\orcidlink{0009-0006-8675-7849}}
\email{fernandobelcks7@gmail.com}
\affiliation{Departamento de Física, Universidade Federal da Paraíba, Centro de Ciências Exatas e da Natureza, 58051-970, João Pessoa, Paraíba, Brazil}

\author{Faizuddin Ahmed\orcidlink{0000-0003-2196-9622}}
\email{faizuddinahmed15@gmail.com}
\affiliation{Department of Physics, The Assam Royal Global University, Guwahati-781035, Assam, India}

\author{Edilberto O. Silva\orcidlink{0000-0002-0297-5747}}
\email{edilberto.silva@ufma.br (Corresponding author)}
\affiliation{Programa de P\'os-Gradua\c c\~ao em F\'{\i}sica \& Coordena\c c\~ao do Curso de F\'isica -- Bacharelado, Universidade Federal do Maranh\~{a}o, 65085-580 S\~{a}o Lu\'is, Maranh\~{a}o, Brazil}

\begin{abstract}
We investigate the propagation of massless fields with spins $s=0$, $s=1$, and $s=1/2$ in the spacetime of an Euler-Heisenberg (EH) black hole surrounded by perfect fluid dark matter (PFDM). This background incorporates both the nonlinear electrodynamic correction associated with the EH effective theory and the logarithmic contribution induced by the surrounding dark matter distribution. After deriving the corresponding Schrödinger-like radial equations, we construct the effective potentials for scalar, electromagnetic, and Dirac perturbations and analyze how they are modified by the black hole charge, the EH parameter, and the PFDM parameter. The greybody factors are estimated through the rigorous Boonserm-Visser lower-bound method, and the associated partial absorption cross sections are obtained for different spin sectors using these bounds. Our results show that the nonlinear electrodynamic and dark matter parameters significantly deform the effective potential barrier, leading to potentially distinguishable changes in the transmission probabilities and absorption spectra within the model. In particular, the greybody lower bounds increase monotonically with the frequency and approach the high-frequency limit, while their low-frequency behavior is strongly affected by the geometry and by the spin of the perturbing field. Moreover, we utilize the greybody lower bound to calculate the energy emission rate. Finally, we make comparisons with the Schwarzschild, Reissner-Nordstr\"om, and pure EH limits, showing that the combined EH+PFDM background produces distinguishable corrections to black-hole scattering within the model. These results highlight greybody bounds as sensitive diagnostic probes of nonlinear electrodynamic effects and dark matter halos around compact objects.
\end{abstract}


\maketitle

\section{Introduction}

Black holes occupy a central position in modern gravitational physics because they provide a natural arena where general relativity, quantum field theory, thermodynamics, and high-energy astrophysics converge. From the classical theory of linear perturbations and quasinormal ringing~\cite{Regge:1957td,Zerilli:1970se,Vishveshwara:1970zz,Teukolsky:1973ha,Chandrasekhar:1975zza,Kokkotas:1999bd,Berti:2009kk,Konoplya:2011qq,Cardoso:2019rvt} to black-hole thermodynamics and phase transitions~\cite{Bekenstein:1973ur,Bardeen:1973gs,Hawking:1975vcx,Hawking:1982dh,Chamblin:1999hg,Chamblin:1999tk,Caldarelli:1999xj,Kastor:2009wy,Dolan:2011xt,Kubiznak:2012wp}, these objects have become indispensable laboratories for testing the consistency of gravitational theories and their possible extensions. In parallel, the direct imaging of compact-object shadows and the detection of gravitational-wave ringdowns have transformed black holes from purely theoretical solutions into observational probes of strong-field gravity~\cite{Synge:1966okc,Falcke:1999pj,EventHorizonTelescope:2019dse,EventHorizonTelescope:2019ths,EventHorizonTelescope:2022wkp,Perlick:2021aok}. These developments motivate the detailed study of modified black-hole backgrounds and of the propagation of test fields around them.

One important class of modifications arises from nonlinear electrodynamics (NLED). The idea that Maxwell electrodynamics may receive nonlinear corrections has a long history, beginning with the Born-Infeld proposal~\cite{Born:1934gh} and the Euler-Heisenberg effective action~\cite{Heisenberg:1936nmg}, later clarified in the context of quantum electrodynamics (QED) by Schwinger's proper-time formalism~\cite{Schwinger:1951nm}. The Euler-Heisenberg (EH) Lagrangian encodes the leading one-loop QED corrections due to virtual electron--positron pairs and predicts nonlinear phenomena such as vacuum birefringence~\cite{BialynickaBirula:1970vy,Novello:1999pg}. When coupled to gravity, nonlinear electromagnetic sources can regularize or deform charged black-hole geometries~\cite{AyonBeato:1998ub,AyonBeato:1999rg,Bronnikov:2000vy,Dymnikova:2004zc} and generate departures from the Reissner-Nordstr\"om solution in the causal structure, thermodynamics, photon motion, and perturbation spectra. In recent years, black holes sourced by nonlinear electromagnetic fields have also been investigated in connection with geodesic motion, optical observables, thermodynamic properties, and perturbative stability~\cite{AlBadawi:2025cjph,Ahmed:2025pduBardeen}.

In the particular case of the Einstein-Euler-Heisenberg system, the first black-hole solutions and their subsequent developments have shown that QED corrections may alter horizon radii, extremality conditions, thermodynamic response functions, and optical properties~\cite{Yajima:2000kw,Ruffini:2013hia,Magos:2020ykt,Guerrero:2020uhn,Amaro:2020xro,Breton:2021mju,Zeng:2022pvb,Dai:2022mko,Li:2022gpf}. These corrections are especially relevant in strong-field regimes, where the electromagnetic invariant can be large enough for nonlinear terms to become dynamically meaningful. In such backgrounds, field propagation is sensitive not only to the metric function itself but also to the way in which NLED modifies the effective potential governing wave scattering. Consequently, EH black holes provide a useful setting for analyzing how quantum-electrodynamic corrections influence transmission probabilities, absorption cross sections, shadows, and quasinormal modes.

A second, independent motivation comes from the possible influence of dark matter on compact-object environments. The existence of dark matter is supported by galactic rotation curves, large-scale structure, and cosmic microwave background measurements~\cite{Zwicky:1933gu,Rubin:1970zza,Navarro:1996gj,Bertone:2004pz,Feng:2010gw,Planck:2018vyg}. Although the microscopic nature of dark matter remains unknown, its macroscopic gravitational effects can be described phenomenologically in several ways. Among these, perfect fluid dark matter (PFDM) provides an analytically tractable model in which the dark sector is represented by an effective fluid distribution around a compact object~\cite{Kiselev:2002dx,Rahaman:2010xs,Li:2012zx,Xu:2017bpz,Zhang:2020mxi}. The corresponding black-hole metrics typically contain logarithmic corrections controlled by a dark matter parameter, which can shift the horizon structure, modify circular orbits, alter the photon sphere, and affect the black hole's optical appearance. Related phenomenological dark-matter environments have recently been analyzed in black-hole spacetimes surrounded by string clouds, monopoles, and effective halo distributions, showing that matter distributions outside the horizon can significantly modify shadows, lensing, thermodynamics, and wave propagation~\cite{Ahmed:2025epjcDM,Ahmed:2025pduPhantom}.

The combination of EH nonlinear electrodynamics and a PFDM environment is therefore physically well-motivated. It allows one to study, within a single analytic background, the simultaneous impact of QED corrections and an effective dark matter halo. Such a geometry was recently constructed for an EH black hole surrounded by PFDM~\cite{Ma:2024psr}. The resulting metric contains both the EH correction proportional to the nonlinear electromagnetic parameter and the logarithmic PFDM contribution. This background has already been explored from the viewpoints of optical properties and thermodynamics, but its perturbative transmission properties and spin-dependent greybody factors remain comparatively underexplored.

Greybody factors are central quantities in black-hole scattering theory. Hawking radiation is thermal only at the horizon. In this case, the spectrum observed at infinity is filtered by the curvature-induced potential barrier surrounding the black hole~\cite{Hawking:1975vcx,Page:1976df,Page:1976ki,Unruh:1976fm,Sanchez:1978qv}. The associated transmission coefficient, or greybody factor, determines the fraction of radiation that escapes to infinity and therefore enters the power spectrum, evaporation rate, and absorption cross section. Over the years, greybody factors have been studied for a wide variety of black-hole geometries and field spins using numerical integration, WKB methods, matching techniques, and rigorous bounds~\cite{Kanti:2004nr,Kanti:2005xa,Boonserm:2008zg,Boonserm:2017qcq,Boonserm:2019mon,Crispino:2013pya,Chowdhury:2020bdi,Zhang:2020qam,Rincon:2020cos,Okyay:2021nnh,Pantig:2022ely,Oshita:2024qnm,Guo:2023nkd}. In particular, the Boonserm-Visser method gives rigorous lower bounds on transmission probabilities in terms of integrals over the effective potential~\cite{Boonserm:2008zg}, making it especially useful when the exact scattering problem cannot be solved analytically. Recent related analyses have also explored greybody factors, scalar-field scattering, quasinormal behaviour, shadows, and perturbative observables in modified black-hole backgrounds involving Lorentz-symmetry violation, nonlinear electrodynamics, global monopoles, quintessence, string clouds, and dark-matter halos~\cite{Belchior:2025krgm,AlBadawi:2022grg,AlBadawi:2025cjph,Ahmed:2025pduBardeen,Ahmed:2025pduPhantom,Ahmed:2025epjcDM}.

The purpose of the present work is to investigate the propagation of massless fields with spins $s=0$, $s=1$, and $s=1/2$ in the background of an EH black hole surrounded by PFDM. We derive the corresponding Schrödinger-like equations, construct the effective potentials, and analyze how the EH and PFDM parameters affect the greybody factors and partial absorption cross sections. Particular attention is paid to the dependence of the transmission probability on the electromagnetic charge, the EH parameter, the PFDM parameter, and the angular quantum numbers. Since each spin sector probes a different effective potential, the comparison among scalar, electromagnetic, and Dirac perturbations provides a useful diagnostic of how nonlinear electrodynamics and dark matter jointly deform the scattering barrier.

This paper is organized as follows. In Sec.~\ref{s2}, we review the EH black hole surrounded by PFDM and discuss the metric function, horizon structure, and relevant parameter ranges. In Sec.~\ref{s3}, we derive the perturbation equations for scalar, electromagnetic, and Dirac fields and reduce them to Schr\"odinger-like form through the tortoise coordinate. In Sec.~\ref{s4}, we compute rigorous lower bounds for the greybody factors by means of the Boonserm-Visser method and obtain the associated absorption cross sections. In Sec.~\ref{s5}, we summarize our results and discuss possible extensions. Throughout this work, we use geometrized units with $G=c=1$ and set $\hbar=1$ when discussing quantum emission quantities.

\section{The Euler-Heisenberg Black Hole Surrounded by Perfect Fluid Dark Matter}\label{s2}

In this section, we present the EH black hole surrounded by PFDM. Such a solution is a static, spherically symmetric solution in Einstein gravity coupled to nonlinear electrodynamics and a non-interacting perfect-fluid dark-matter halo. This model, originally constructed by Ma \emph{et al.}~\cite{Ma:2024psr}, superposes the leading-order perturbative EH correction (arising from one-loop QED vacuum polarization) with a phenomenological PFDM component whose energy density scales as $\rho_{\text{PFDM}} \propto 1/r^3$. The solution is analytically tractable and reduces to the pure EH, Reissner-Nordstr\"{o}m (RN), and RN plus PFDM limits, and allows us to study the connection between quantum electrodynamic effects and dark matter halos in strong-gravity regimes. In this sense, as a starting point, we consider the total gravitational action given by
\begin{align}
S = \int d^4x\,\sqrt{-g}\left[\frac{R}{16\pi} + \sum_i\mathcal{L}^{(i)}\right],
\end{align}
where $\mathcal{L}^{(i)}$ denotes the Lagrangian density of each uncoupled matter sector, namely EH nonlinear electrodynamics, PFDM, and the cosmological-constant term treated as an additional fluid. Then, the Einstein equations read
\begin{align}
R_{\mu\nu} - \frac{1}{2}Rg_{\mu\nu} = 8\pi T_{\mu\nu},
\end{align}
with
\begin{align}
T_{\mu\nu} = \sum_i T_{\mu\nu}^{(i)}
\end{align}
Because the sectors do not interact, the metric function can be obtained by linear superposition. For a static, spherically symmetric line element
\begin{align}
ds^2 = -g(r)\,dt^2 + \frac{dr^2}{g(r)} + r^2\,d\Omega^2,
\end{align}
the function $g(r)$ satisfies
\begin{align}
g(r) = 1 -\frac{2M}{r}+ \sum_i f_i(r),
\end{align}
where each $f_i(r)$ is determined independently from the corresponding energy-momentum component $T_{\mu\nu}^{(i)}$. The EH sector with perturbative expansion in the small QED parameter $a > 0$ contributes with the following form
\begin{align}
f_{\text{EH}}(r) = \frac{Q^2}{r^2} - \frac{aQ^4}{20r^6}.
\end{align}
On the other hand, the PFDM sector has the following stress-energy component
\begin{align}
T_t^t = \frac{\alpha}{8\pi r^3},
\end{align}
where $\alpha$ is the dimensionless dark-matter intensity parameter. The PFDM yields the following logarithmic correction
\begin{align}
f_{\text{PFDM}}(r) = \frac{\alpha}{r}\ln\left(\frac{r}{|\alpha|}\right).
\end{align}
In anti-de Sitter (AdS) spacetime, the cosmological-constant term adds $-\Lambda r^2/3$ ($\Lambda < 0$). The complete metric function is therefore
\begin{equation}
g(r) = 1 - \frac{2M}{r} + \frac{Q^2}{r^2} - \frac{aQ^4}{20r^6} + \frac{\alpha}{r}\ln\left(\frac{r}{|\alpha|}\right) - \frac{\Lambda r^2}{3}.
\label{eq:metric}
\end{equation}
Here, $M$ is the ADM mass, $Q$ the electric charge, and $a = 8\alpha_{\text{EM}}^2/(45m_e^4)$ encodes the strength of vacuum polarization ($\alpha_{\text{EM}}$ is the fine-structure constant). The solution is asymptotically flat when $\Lambda = 0$ and asymptotically AdS otherwise. The curvature singularity remains at $r = 0$; the event horizon(s) $r_h$ are the largest positive root(s) of $g(r_h) = 0$.

As discussed in this section, the EH black hole solution with PFDM halo provides a simple realization of a $1/r^3$ density profile inspired by galactic rotation curves, while the EH term encodes realistic QED vacuum polarization, which is tiny in Nature, $a\sim 10^{-30}$ in Planck units. Besides, the uncoupled superposition guarantees consistency at the classical level. Astrophysical implications include modified quasi-periodic oscillations, ringdown frequencies, and energy-collision processes near the horizon. It is worth highlighting that some extensions already explored in the literature include simultaneous quintessence dark energy, rotating (Kerr-like) generalizations, plasma lensing in weak plasma fields, and observational constraints from Event Horizon Telescope shadow data.

\begin{figure*}[ht!]
\begin{center}
\includegraphics[scale=0.55]{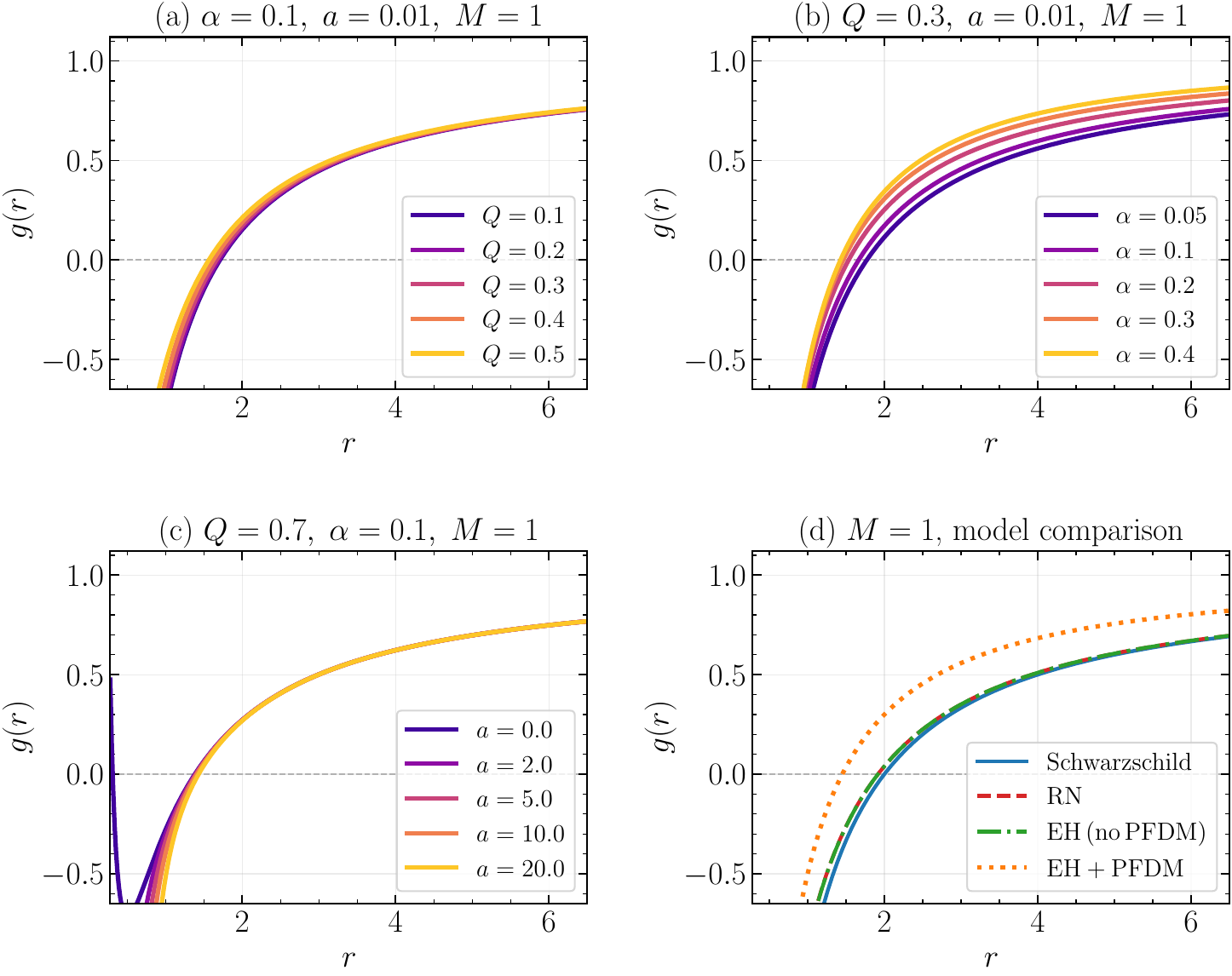}
\end{center}
\vspace{-0.4cm}
\caption{Metric function $g(r)$ for $M=1$ and $\Lambda=0$.
(a)~Varying $Q\in\{0.1,0.2,0.3,0.4,0.5\}$, fixed $\alpha=0.1$, $a=0.01$.
(b)~Varying $\alpha\in\{0.05,0.10,0.20,0.30,0.40\}$, fixed $Q=0.3$, $a=0.01$.
(c)~Varying $a\in\{0,2,5,10,20\}$, fixed $Q=0.7$, $\alpha=0.1$; the range is illustrative ($a_{\rm phys}\sim 10^{-30}$).
(d)~Model comparison: Schwarzschild ($Q=\alpha=a=0$), RN ($Q=0.4$, $\alpha=a=0$), EH ($Q=0.4$, $a=5$, $\alpha=0$), and EH+PFDM ($Q=0.4$, $a=5$, $\alpha=0.25$). Dashed horizontal line marks $g=0$.}
\label{fig:metric}
\end{figure*}

Figure~\ref{fig:metric} illustrates the structure of the metric function $g(r)$ for representative values of the model parameters with $M = 1$ and $\Lambda = 0$. Panel~(a) shows the effect of increasing the electric charge $Q$: the horizon radius $r_h$ (zero-crossing of $g$) decreases as $Q$ grows, reflecting the strengthening of the electromagnetic repulsion near the singularity. Panel~(b) demonstrates the role of the PFDM parameter $\alpha$: the logarithmic term $(\alpha/r)\ln(r/|\alpha|)$ is positive for $\alpha > 0$ and $r > |\alpha|$, so it increases $g(r)$ and shifts the zero-crossing to smaller radii --- a larger $\alpha$ produces a smaller $r_h$ and therefore a stronger filtering of Hawking radiation. Panel~(c) isolates the influence of the EH nonlinearity parameter $a$: the correction $-aQ^4/(20r^6)$ is negative, reducing $g(r)$ near the singularity and pushing the horizon outward --- \emph{opposite} to the effects of $Q$ and $\alpha$. For astrophysically motivated values ($a\sim 10^{-30}$ in Planck units), this correction is negligible; the range shown in panel~(c) is chosen for illustrative purposes. Finally, panel~(d) compares four representative models: Schwarzschild ($Q=\alpha=a=0$), Reissner-Nordstr\"{o}m (RN, $\alpha=a=0$), EH without PFDM ($\alpha=0$), and the full EH+PFDM solution. The RN and EH curves are visually indistinguishable for moderate values of $a$, confirming that the PFDM logarithmic term is the dominant modifier of the metric with respect to the RN baseline.

\section{External perturbations and effective potential}\label{s3}

In this section, we will examine external perturbations and derive the effective potentials for spin 0, spin 1, and spin $1/2$. As a starting point, one considers a generic static, spherically symmetric black hole spacetime described by the following line element
\begin{align}
ds^{2} = -f(r)\,dt^{2} + \frac{dr^{2}}{f(r)} + r^{2}\left(d\theta^{2} + \sin^{2}\theta\,d\phi^{2}\right),
\label{eq:generic-metric}
\end{align}
where the metric function $f(r)$ satisfies the usual asymptotic flatness conditions $f(r\to\infty)=1$ and $f(r_{h})=0$ at the event horizon(s) $r_{h}$. The tortoise coordinate $r_{*}$ is defined via
\begin{align}
dr_{*} = \frac{dr}{f(r)},
\end{align}
or
\begin{align}
\frac{d}{dr_{*}} = f(r)\frac{d}{dr}.
\end{align}
For massless test fields, we seek time-harmonic solutions of the form $e^{-i\omega t}$, with $\omega$ being the frequency, and separate the angular dependence using the appropriate spherical (or spin-weighted) harmonics. After substitution into the field equations, the radial dynamics reduce to a one-dimensional Schr\"{o}dinger-like wave equation, namely
\begin{align}
\frac{d^{2}\psi}{dr_{*}^{2}} + \bigl(\omega^{2} - V(r)\bigr)\psi = 0,
\end{align}
where $\psi(r)$ is a suitably rescaled radial wave function and $V(r)$ is the spin-dependent effective potential. Below we derive $V(r)$ explicitly for spin-0 (scalar), spin-1 (electromagnetic), and spin-1/2 (Dirac) fields. All expressions are valid for an arbitrary $f(r)$, e.g., Schwarzschild, Reissner-Nordstr\"{o}m, Euler-Heisenberg, or Euler-Heisenberg along with PFDM.

\subsection{Spin-0: Massless Scalar Field (Klein--Gordon Equation)}

First, let us obtain the effective potential for massless spin-0 particles. In this case, the massless scalar field $\Phi$ obeys the Klein-Gordon equation given by
\begin{align}
\Box\Phi \equiv \frac{1}{\sqrt{-g}}\partial_{\mu}\Bigl(\sqrt{-g}\,g^{\mu\nu}\partial_{\nu}\Phi\Bigr) = 0.
\end{align}
In the metric \eqref{eq:generic-metric} one has $\sqrt{-g}=r^{2}\sin\theta$. We separate variables as follows
\begin{align}
\Phi(t,r,\theta,\phi) = \frac{1}{r}\psi(r)\,e^{-i\omega t}\,Y_{lm}(\theta,\phi),
\end{align}
where $Y_{lm}$ are the spherical harmonics satisfying $\Delta_{S^{2}}Y_{lm}=-l(l+1)Y_{lm}$. Substituting into the Klein-Gordon equation and collecting the radial part yields the second-order ordinary differential equation, namely
\begin{align}
\frac{d}{dr}\Bigl(f\,r^{2}\frac{dR}{dr}\Bigr) + \left(\frac{\omega^{2}r^{2}}{f}-l(l+1)\right)R=0,
\end{align}
with $\Phi=R(r)Y_{lm}e^{-i\omega t}$. Defining the rescaled radial function $\psi(r)=rR(r)$ and changing to the tortoise coordinate $r_{*}$ (so that $\frac{d}{dr}=f\frac{d}{dr_{*}}$) produces, after a straightforward but lengthy differentiation,
\begin{align}
\frac{d^{2}\psi}{dr_{*}^{2}} + \Bigl[\omega^{2} - V_{0}(r)\Bigr]\psi = 0,
\end{align}
where the effective potential reads
\begin{equation}
V_{0}(r) = f(r)\left[\frac{l(l+1)}{r^{2}} + \frac{f'(r)}{r}\right].
\label{eq:V0}
\end{equation}
The term proportional to $f'(r)$ arises from the radial derivative acting on the metric function when the Laplacian is evaluated in curved spacetime.

\subsection{Spin-1: massless electromagnetic field}

Now, let us derive the effective potential for a spin-1 field. In this case, the source-free Maxwell equations read
\begin{align}
\nabla_{\mu}F^{\mu\nu}=0,
\end{align}
where $F_{\mu\nu}=\nabla_{\mu}A_{\nu}-\nabla_{\nu}A_{\mu}$ is the strength field. To proceed further, we will work in the Lorenz gauge $\nabla_{\mu}A^{\mu}=0$. For a static spherically symmetric background, the electromagnetic perturbations can be decomposed into odd-parity (axial) and even-parity (polar) sectors using vector spherical harmonics. Both sectors decouple and, remarkably, obey identical radial equations (isospectrality). Introducing the gauge-invariant radial function $\psi(r)$ (a linear combination of the vector-potential components) and performing the following separation
\begin{align}
A_{\mu}(t,r,\theta,\phi)=e^{-i\omega t}\,Y_{lm}^{(\text{vector})}(\theta,\phi)\,\frac{\psi(r)}{r},
\end{align}
the wave equation again reduces to the Schr\"{o}dinger form, namely
\begin{align}
\frac{d^{2}\psi}{dr_{*}^{2}} + \Bigl[\omega^{2} - V_{1}(r)\Bigr]\psi = 0.
\end{align}
The effective potential for spin-1 is given by
\begin{equation}
V_{1}(r) = f(r)\frac{l(l+1)}{r^{2}},
\label{eq:V1}
\end{equation}
with $l\geq1$. Note the complete absence of the $f'(r)$ term; the extra curvature contribution present for scalars cancels exactly for vectors because of the antisymmetric structure of $F_{\mu\nu}$. This potential governs photon propagation, quasinormal ringing, and grey-body factors of the black hole.

\subsection{Spin-1/2: Massless Dirac Field}

The massless Dirac field $\Psi$ satisfies the Dirac equation in curved spacetime, namely
\begin{align}
i\gamma^{\mu}\nabla_{\mu}\Psi=0,
\end{align}
where $\nabla_{\mu}=\partial_{\mu}+\frac{1}{4}\omega_{\mu ab}\sigma^{ab}$ includes the spin connection $\omega_{\mu ab}$ built from the tetrad formalism. We will adopt the following orthonormal tetrad
\begin{align}
& e^{\hat{0}}=\sqrt{f}\,dt,\\
& e^{\hat{1}}=\frac{dr}{\sqrt{f}},\\
& e^{\hat{2}}=r\,d\theta,\\
& e^{\hat{3}}=r\sin\theta\,d\phi.
\end{align}
The gamma matrices satisfy the usual Clifford algebra in the tetrad basis, and the spin connection coefficients are computed from the Cartan structure equations. After a standard separation of variables using the spinor spherical harmonics $\chi_{\kappa m}(\theta,\phi)$ (with $\kappa=\pm(l+1)$, $l=0,1,2,\dots$), the four-component Dirac spinor reduces to a pair of radial functions $u(r_{*})$ and $v(r_{*})$ obeying the following first-order coupled system
\begin{align}
&\frac{du}{dr_{*}} = (\omega - W)v,\\
&\frac{dv}{dr_{*}} = -(\omega + W)u,
\end{align}
where the superpotential $W(r)$ reads
\begin{align}
W(r)=\frac{\kappa}{r}\sqrt{f(r)}.
\end{align}
Eliminating one function (e.g., $v$) in favor of the other yields two decoupled second-order Schr\"{o}dinger equations, one for each sign given by
\begin{align}
\frac{d^{2}\psi_{\pm}}{dr_{*}^{2}} + \bigl(\omega^{2} - V_{\pm}(r)\bigr)\psi_{\pm}=0.
\end{align}
The effective potentials are obtained from the supersymmetric quantum mechanics identity, namely
\begin{align}
V_{\pm}(r)=W^{2}\pm\frac{dW}{dr_{*}}.
\end{align}
Explicitly, this potential reads
\begin{equation}
V_{\pm}(r)=\frac{\kappa}{r^2}\bigg(\kappa f\pm\frac{rf^{\prime}}{2}\sqrt{f}\mp f^{3/2}\bigg).
\label{eq:Vpm}
\end{equation}
Note that the derivative $\frac{dW}{dr_{*}}=f\,dW/dr$ produces the linear-in-$f'$ correction. The two potentials $V_{+}$ and $V_{-}$ correspond to the two chiral sectors. They are isospectral in the sense that they produce identical quasinormal-mode spectra for the massless Dirac field on a static black-hole background.

The three potentials can be written compactly using the spin $s$ as follows
\begin{align}
V_{s}(r)=f(r)\left[\frac{l(l+1)}{r^{2}} + (1-s^{2})\frac{f'(r)}{r}\right].
\end{align}
Above, for bosons ($s=0,1$), we recover the scalar and vector potentials derived before. Besides, note that $s=2$ gravitational perturbations recover the famous Regge-Wheeler/Zerilli form. For fermions ($s=1/2$) the supersymmetric structure \eqref{eq:Vpm} replaces the simple bosonic expression. All potentials vanish at spatial infinity ($V\to0$ as $r\to\infty$) and at the horizon ($V(r_{h})=0$), guaranteeing the correct boundary conditions for quasinormal modes ($\psi\sim e^{\pm i\omega r_{*}}$ at $r_{*}\to\pm\infty$).

These effective potentials are directly applicable to any black-hole solution of the form \eqref{eq:generic-metric}, including the EH black hole and its embedding in PFDM discussed previously. They govern the computation of quasinormal frequencies (via continued-fraction or shooting methods), grey-body factors, superradiance, and late-time power-law tails, providing a unified framework for probing the stability and ringdown signatures of quantum-corrected and dark-matter-surrounded black holes.

\begin{figure*}[ht!]
\begin{center}
\includegraphics[scale=0.55]{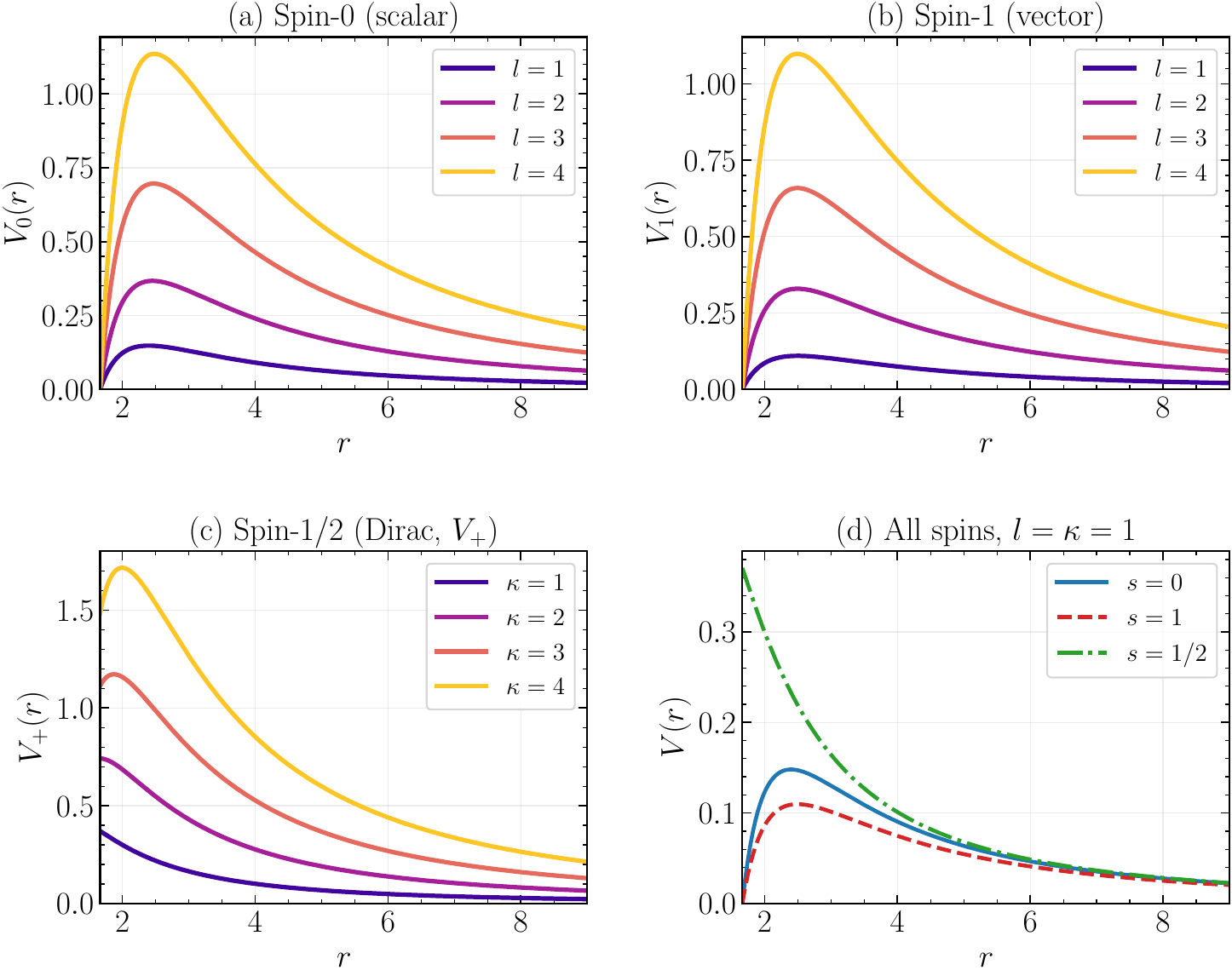}
\end{center}
\vspace{-0.4cm}
\caption{Effective potentials for $M=1$, $Q=0.3$, $a=1.0$, $\alpha=0.1$, plotted from the horizon $r_h$ outward.
(a)~Scalar potential $V_0(r)$ for $l=1,2,3,4$.
(b)~Vector potential $V_1(r)$ for $l=1,2,3,4$.
(c)~Dirac potential $V_+(r)$ for $\kappa=1,2,3,4$.
(d)~Comparison of all three spins for $l=\kappa=1$, showing the hierarchy $V_0>V_1>V_+$.}
\label{fig:potentials}
\end{figure*}

Figure~\ref{fig:potentials} shows the three effective potentials for reference parameters $M=1$, $Q=0.3$, $a=1.0$, $\alpha=0.1$. Panels~(a) and~(b) display the scalar $V_0(r)$ and vector $V_1(r)$ potentials for angular quantum numbers $l=1,2,3,4$; increasing $l$ raises the barrier height and shifts its peak inward, consistent with the centrifugal-barrier analogy. Panel~(c) shows the Dirac potential $V_+(r)$ for $\kappa=1,2,3,4$. Panel~(d) compares all three spins for $l=\kappa=1$, revealing the hierarchy $V_0 > V_1 > V_+$ at the peak: the scalar potential is the tallest owing to the curvature-coupling term $f'(r)/r$ in Eq.~\eqref{eq:V0}, which is absent for the vector field (Eq.~\eqref{eq:V1}) and is replaced by the supersymmetric structure for the Dirac field (Eq.~\eqref{eq:Vpm}). This ordering is directly responsible for the transmission hierarchy $T_{1/2}>T_1>T_0$ observed for the representative configurations discussed in Section~\ref{s4}.

\section{Rigorous greybody bounds for test fields}\label{s4}

After deriving the effective potential for test fields, the next step is to examine greybody factors and absorption for these fields. In this context, one knows that the spectrum of Hawking radiation emitted by a black hole is not purely thermal. As particles propagate from the near-horizon region to spatial infinity, the nontrivial spacetime curvature scatters part of the radiation back into the black hole. The frequency- and spin-dependent factor that quantifies the fraction of radiation that reaches a distant observer is called the greybody factor or transmission coefficient. This section provides a rigorous derivation of lower bounds for the greybody factors of massless test fields of spin $s = 0$ (scalar), $s = 1$ (electromagnetic), and $s = 1/2$ (Dirac) propagating in a generic static spherically symmetric black-hole spacetime.

The Hawking temperature is $T_{H} = f'(r_{h})/(4\pi)$. The differential emission rate for a field of spin $s$ and angular momentum quantum numbers $(l,m)$ is modified from the pure Planckian spectrum
\begin{align}
\frac{d^{2}N_{s}}{d\omega\,dt} = \frac{1}{2\pi}\frac{T_{s}(\omega,l)}{e^{\omega/T_{H}} \pm 1}\,d\omega\,dt,
\end{align}
where the sign is $+$ for fermions and $-$ for bosons, and $T_{s}(\omega,l)$ is the greybody factor (averaged over $m$ if necessary). In the high-frequency geometric-optics limit $\omega \gg T_{H}$, $\Gamma_{s} \to 1$. At low frequencies, $\Gamma_{s}$ is strongly suppressed, making the spectrum ``grey'' rather than black.

The greybody factor is rigorously identified with the transmission probability through the effective potential barrier as follows
\begin{align}
T_{s}(\omega,l) = |T(\omega,l)|^{2} = 1 - |R(\omega,l)|^{2},
\end{align}
where $T$ and $R$ are the transmission and reflection coefficients for scattering from infinity. From the effective potentials $V_{s}(r)$ derived in the previous section (for spin 0, 1, and $\pm$ for spin $1/2$), the radial wave equation in tortoise coordinates $r_{*}$ (with $dr_{*} = dr/f(r)$) reads
\begin{align}
\frac{d^{2}\psi}{dr_{*}^{2}} + \bigl(\omega^{2} - V_{s}(r)\bigr)\psi = 0.
\end{align}
The boundary conditions for a wave incident from infinity are:
- As $r_{*} \to +\infty$ ($r \to \infty$): incoming wave plus reflected wave, we write
\begin{align}
\psi(r_{*}) \sim e^{-i\omega r_{*}} + R(\omega)\,e^{+i\omega r_{*}}.
\end{align}
- As $r_{*} \to -\infty$ ($r \to r_{h}^{+}$): purely ingoing wave (no outgoing flux from the horizon), we write
\begin{align}
  \psi(r_{*}) \sim T(\omega)\,e^{-i\omega r_{*}}.
\end{align}

The greybody factor is then $T_s(\omega) = |T(\omega)|^{2}$. For each spin, the potential $V_{s}(r)$ is different, leading to spin-dependent transmission. At this stage, it is important to point out that the exact greybody factor $\Gamma_s(\omega,l)$ for Hawking radiation of spin-$s$ fields cannot, in general, be computed exactly for realistic black-hole metrics. A powerful model-independent approach, introduced by Boonserm and Visser (2008) \cite{Boonserm:2008zg} and systematically developed in subsequent works (Boonserm, Ngampitipan, Visser, and collaborators), provides rigorous lower bounds on $\Gamma_s$ that depend only on the shape of the effective potential $V_s(r)$ and the frequency $\omega$. These bounds are derived from the one-dimensional scattering problem in quantum mechanics and hold for any potential that vanishes at the horizon and at infinity, is positive in between, and satisfies mild integrability conditions. They are particularly useful for comparing different black-hole models, such as Schwarzschild, Reissner-Nordstr\"{o}m, EH, or EH plus PFDM, without solving the full wave equation numerically.

As derived in the previous section, the radial dynamics of a massless test field of spin $s$ reduce to the Schrödinger-type equation in tortoise coordinates $r_*$ ($dr_* = dr/f(r)$)
\begin{align}
\frac{d^2\psi}{dr_*^2} + \bigl(\omega^2 - V_s(r)\bigr)\psi = 0,
\end{align}
where $V_s(r)$ is the effective potential for spin $s = 0$, $s = 1$, or the pair $V_\pm$ for $s = 1/2$ (explicit forms given earlier). The greybody factor is
\begin{align}
T_s(\omega,l) = |T(\omega,l)|^2 = 1 - |R(\omega,l)|^2,
\end{align}
with boundary conditions of a purely ingoing wave at the horizon ($r_* \to -\infty$) and an incident wave from infinity ($r_* \to +\infty$).
The bound follows from the conservation of the Wronskian of the scattering solutions and the use of a $2 \times 2$ transfer-matrix formalism (or, equivalently, from monotonicity arguments in the phase-space formulation of the Schr\"{o}dinger equation). Consider the general transmission coefficient for a potential barrier $V(r_*)$ that satisfies $V(r_* \to \pm\infty) = 0$ and $V(r_*) \geq 0$, thereby leading to
\begin{align}
 \int_{-\infty}^{+\infty} |V(r_*)|\,dr_* < \infty.
\end{align}
In this way, Boonserm and Visser proved that the transmission probability obeys the following exact inequality
\begin{equation}
|T(\omega)|^2 \geq \mathrm{sech}^2\left(\frac{1}{2\omega}\int_{-\infty}^{+\infty}|V(r_*)|\,dr_*\right).
\label{eq:boonserm-visser-bound}
\end{equation}
Equivalently, after restoring the radial coordinate, the above expression takes the form
\begin{equation}
T_s(\omega) \geq \mathrm{sech}^2\left(\frac{1}{2\omega}\int_{r_h}^{\infty}\frac{|V(r)|}{f(r)}\,dr\right).
\label{eq:radial-bound}
\end{equation}
This bound is saturated in the low-frequency limit for certain exactly solvable potentials and becomes increasingly tight for high multipole orders $l$ or high frequencies, where the potential barrier is narrow. An analogous upper bound on the reflection coefficient is given by
\begin{align}
|R(\omega)|^2 \leq \tanh^2\left(\frac{1}{2\omega}\int_{r_h}^{\infty}\frac{|V(r)|}{f(r)}\,dr\right).
\end{align}
For fermions ($s=1/2$), the supersymmetric structure of the pair $V_\pm$ allows one to apply the bound separately to each chiral sector and then combine the results. The bound remains of the same functional form but is evaluated on the appropriate superpotential-derived potential.
For the scalar ($s=0$) and electromagnetic ($s=1$) cases, substitute the following effective potentials
\begin{align}
&V_0(r) = f(r)\left[\frac{l(l+1)}{r^2} + \frac{f'(r)}{r}\right],\\ &V_1(r) = f(r)\frac{l(l+1)}{r^2}.
\end{align}
The integral in \eqref{eq:radial-bound} then becomes
\begin{align}
I = \int_{r_h}^{\infty}\frac{|V_s(r)|}{f(r)}\,dr = \int_{r_h}^{\infty}\left[\frac{l(l+1)}{r^2} + (1-s^2)\frac{f'(r)}{r}\right]dr,
\end{align}
which can be evaluated analytically for many models (the $f'$ term telescopes). For the Dirac case, the integral is performed over $|V_\pm(r)|$ expressed in terms of the superpotential $W(r) = \kappa\sqrt{f(r)}/r$.

In the Euler-Heisenberg black hole (with or without perfect-fluid dark matter), the metric function $f(r)$ (or $g(r)$) contains the extra terms $-\frac{aQ^4}{20r^6}$ and $+\frac{\alpha}{r}\ln(r/|\alpha|)$. These corrections modify $V_s(r)$ only mildly at large $r$ but can alter the integral $I$ near the horizon, tightening or loosening the lower bound on $\Gamma$. Because the bound depends solely on the integrated strength of the potential barrier, one can compare the EH + PFDM model directly with the Reissner-Nordstr\"{o}m case by computing $I$ numerically for given parameters $(M,Q,a,\alpha)$.

The original bound can be sharpened by replacing $V$ with a monotonic majorant or by using a family of test functions $h(r_*)$ in the transfer-matrix formalism, yielding
\begin{align}
|T|^2 \geq \mathrm{sech}^2\left(\frac{1}{2\omega}\int_{-\infty}^{+\infty}h(r_*)\,|U(r_*)|\,dr_*\right),
\end{align}
where $U$ is a normalized potential function. In the high-$l$ and high-$\omega$ limits, the bound approaches the geometric-optics value $\Gamma\to 1$ exponentially fast. In the case of massive fields, the formalism extends straightforwardly by adding a mass term to the effective potential. Finally, for rotating black hole solutions, such as Kerr-like metrics, the bound generalizes to the Teukolsky equation with frequency-dependent superradiant terms, but the core integral structure remains similar.

In this sense, the rigorous lower bound provides a conservative estimate of the Hawking flux without requiring full numerical integration of the wave equation. For the EH+PFDM black hole, the PFDM logarithmic term typically increases the integrated barrier height, thereby decreasing the lower bound on $\Gamma$ at low frequencies (stronger suppression of emission), while the small QED parameter $a$ produces only perturbative corrections. These bounds are therefore useful for probing dark-matter halo parameters $\alpha$ via hypothetical observations of Hawking radiation (e.g., in analog systems or primordial black holes) and for proving stability of the evaporation process.

Therefore, the Boonserm-Visser bound \eqref{eq:radial-bound} furnishes a rigorous and model-independent lower limit on greybody factors that applies uniformly to any static spherically symmetric black hole, including the quantum-corrected EH solutions discussed earlier. It closes the gap between analytic insight and numerical phenomenology in black-hole thermodynamics.

Before presenting the results for each spin sector, we state the explicit formula for the absorption cross section used throughout this section. For a mode with angular momentum quantum number $l$ (or $\kappa$ for the Dirac case), the partial-wave absorption cross section is
\begin{equation}
\sigma_s(\omega, l) = \frac{\pi(2l+1)}{\omega^2}\,T_s(\omega, l).
\label{eq:sigma}
\end{equation}
The factor $(2l+1)/\omega^2$ combines the angular-momentum degeneracy with the geometric phase space for the partial modes considered here. For fields with spin, and especially for the Dirac sector, this expression should be understood as the partial-mode prescription used in the present comparison, with the angular labels interpreted according to the corresponding spin harmonics. Because $\Gamma_s \to 0$ at low $\omega$ faster than $\omega^2 \to 0$ for the nonzero angular modes shown in the figures, one has $\sigma_s \to 0$ for $\omega \to 0$ in those channels; conversely, at high $\omega$ the prefactor $1/\omega^2$ drives $\sigma_s \to 0$ even though $\Gamma_s \to 1$. This statement refers to the plotted partial waves and should not be confused with the well-known total low-frequency scalar absorption limit associated with the $l=0$ mode. The competition between these two limits produces the characteristic broad peak at intermediate frequencies observed in all panels below.

\subsection{Spin-0}
For the spin-0 case, the Boonserm-Visser lower bound for the greybody factor, written in terms of the horizon radius $r_h$, reads
\begin{align}
    T_{\rm spin\,0}&=\mathrm{sech}^2\!\Bigg[\frac{1}{2\omega}\Bigg(\frac{3aQ^4}{70r_h^7}+\frac{l(l+1)}{r_h}\notag\\&+\frac{105\alpha+420M-210\alpha\ln\!\left(\dfrac{r_h}{\alpha}\right)}{420r_h^2}-\frac{2Q^2}{3r_h^3}\Bigg)\Bigg]
    \label{eq:T0}
\end{align}
In this case, we see that the lower bound $T_{\rm spin\,0}(\omega)$ increases monotonically with the frequency. This behavior is expected: in the low-frequency regime, the wave does not carry enough energy to cross the curvature-induced barrier efficiently, whereas at higher frequencies the barrier becomes progressively less important and the transmission approaches unity. The effect of the charge $Q$ is to reduce the scalar transmission across the entire frequency range analyzed. Physically, increasing the charge strengthens the non-Schwarzschild character of the geometry and reduces the horizon radius, which in turn raises the effective obstruction to the propagation of scalar modes.
The same qualitative tendency is produced by the dark matter parameter $\alpha$. As $\alpha$ grows, the scalar greybody lower bound becomes smaller, indicating that the dark matter environment further suppresses the probability that Hawking quanta escape to infinity.

An interesting feature of the scalar channel is that suppression is not solely determined by the horizon radius. The analytical expression in Eq.~\eqref{eq:T0} shows that the barrier contains explicit corrections involving both $Q$ and $\alpha$. This means that the scalar sector is directly sensitive to the combined effects of nonlinear electrodynamics and the dark matter halo. As a result, the scalar mode is the most strongly filtered channel among the three cases considered here.

For the scalar field, increasing either $Q$ or $\alpha$ lowers the height of this peak and shifts it to slightly larger frequencies. This means that the black hole becomes a less efficient absorber and that the frequency window of maximum absorption moves to the right as the geometry departs further from the Schwarzschild limit. The same pattern is observed in the vector and fermionic channels. In other words, a stronger electric charge and a denser dark matter environment both make it more difficult for the black hole to absorb low- and intermediate-energy quanta. The absorption is not simply reduced; the characteristic frequency scale of the process is also modified.

\begin{figure*}[ht!]
\centering
\begin{minipage}[t]{0.48\textwidth}
    \centering
    \includegraphics[height=4.8cm]{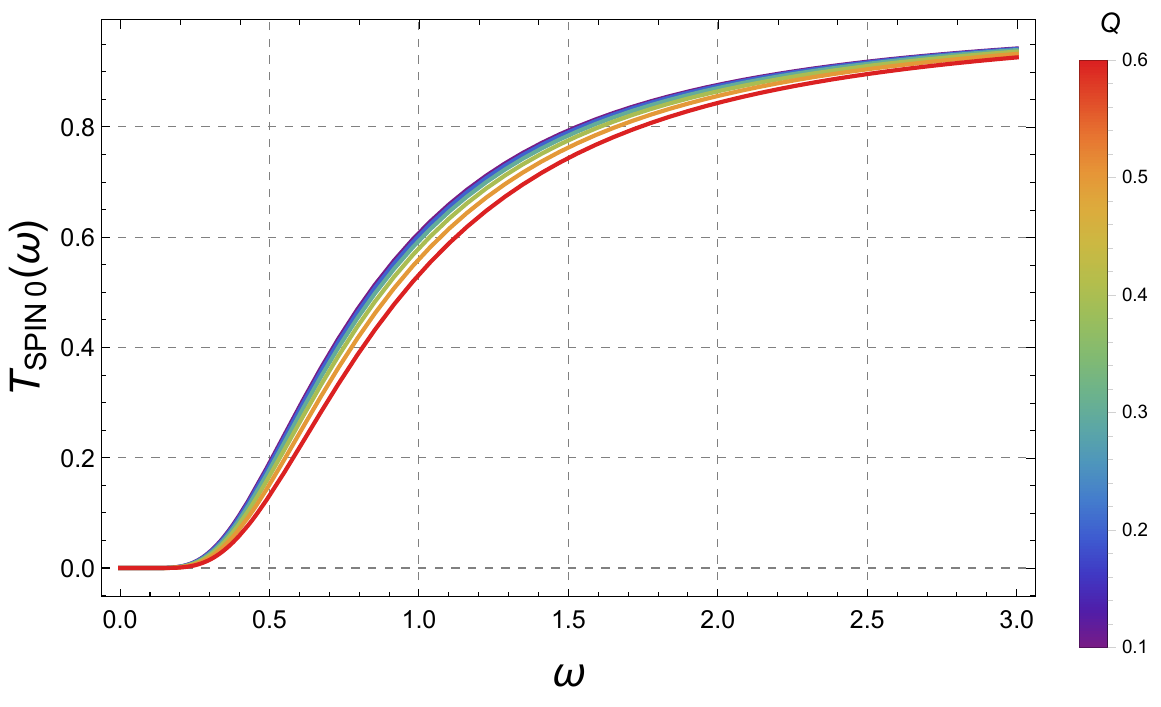}

    \vspace{0.15cm}
    \parbox{0.95\textwidth}{\centering\footnotesize
    (a) $T_0(\omega)$: varying $Q\in\{0.1,\ldots,0.6\}$; 
    $\alpha=0.01,\,l=1,\,M=1$}
\end{minipage}
\hfill
\begin{minipage}[t]{0.48\textwidth}
    \centering
    \includegraphics[height=4.8cm]{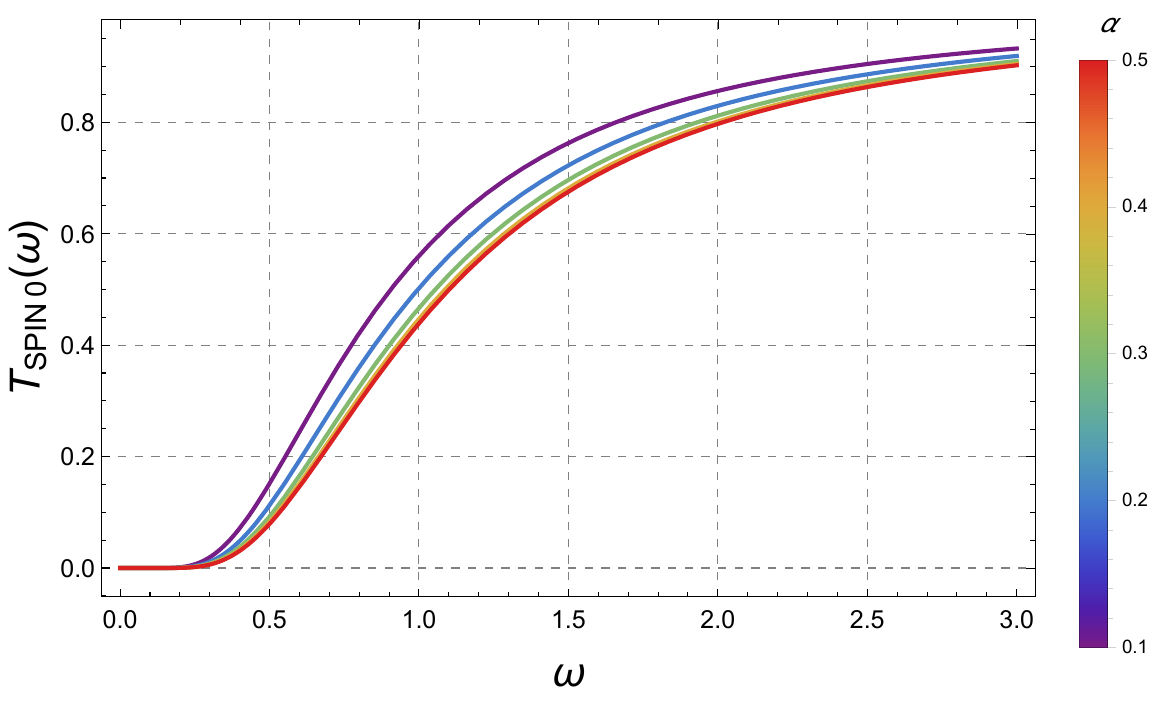}

    \vspace{0.15cm}
    \parbox{0.95\textwidth}{\centering\footnotesize
    (b) $T_0(\omega)$: varying $\alpha\in\{0.1,\ldots,0.5\}$; 
    $Q=0.1,\,l=1,\,M=1$}
\end{minipage}

\vspace{0.45cm}

\begin{minipage}[t]{0.48\textwidth}
    \centering
    \includegraphics[height=4.8cm]{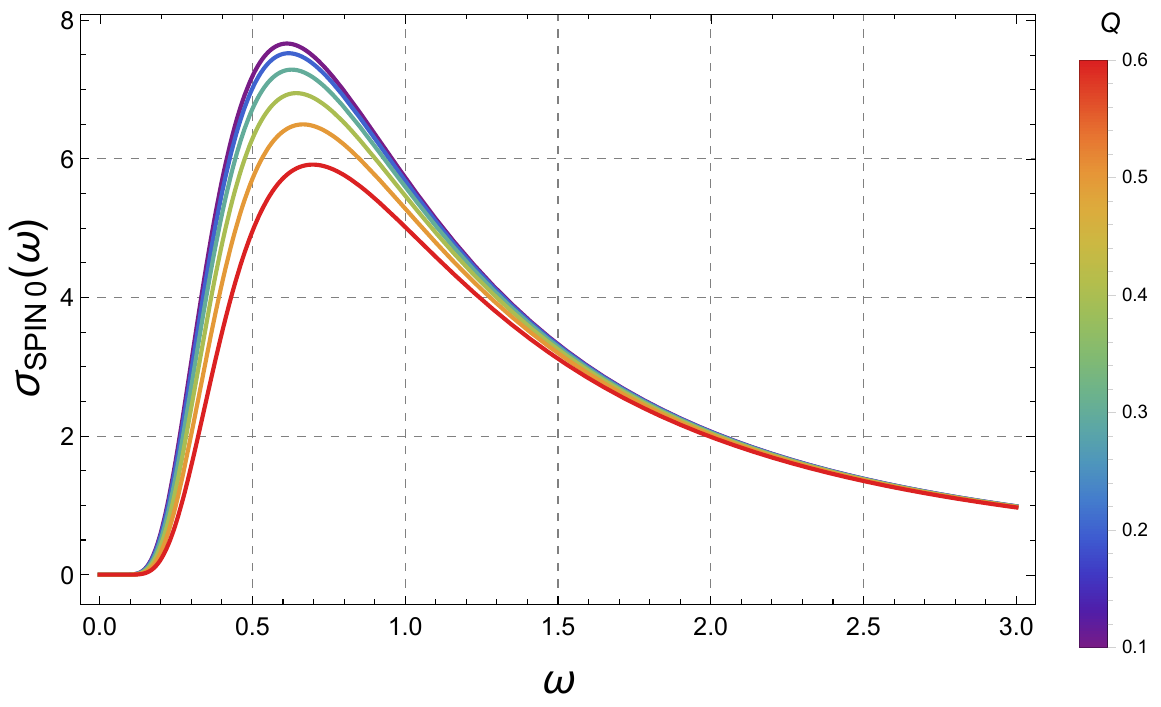}

    \vspace{0.15cm}
    \parbox{0.95\textwidth}{\centering\footnotesize
    (c) $\sigma_0(\omega)$: same parameters as in (a)}
\end{minipage}
\hfill
\begin{minipage}[t]{0.48\textwidth}
    \centering
    \includegraphics[height=4.8cm]{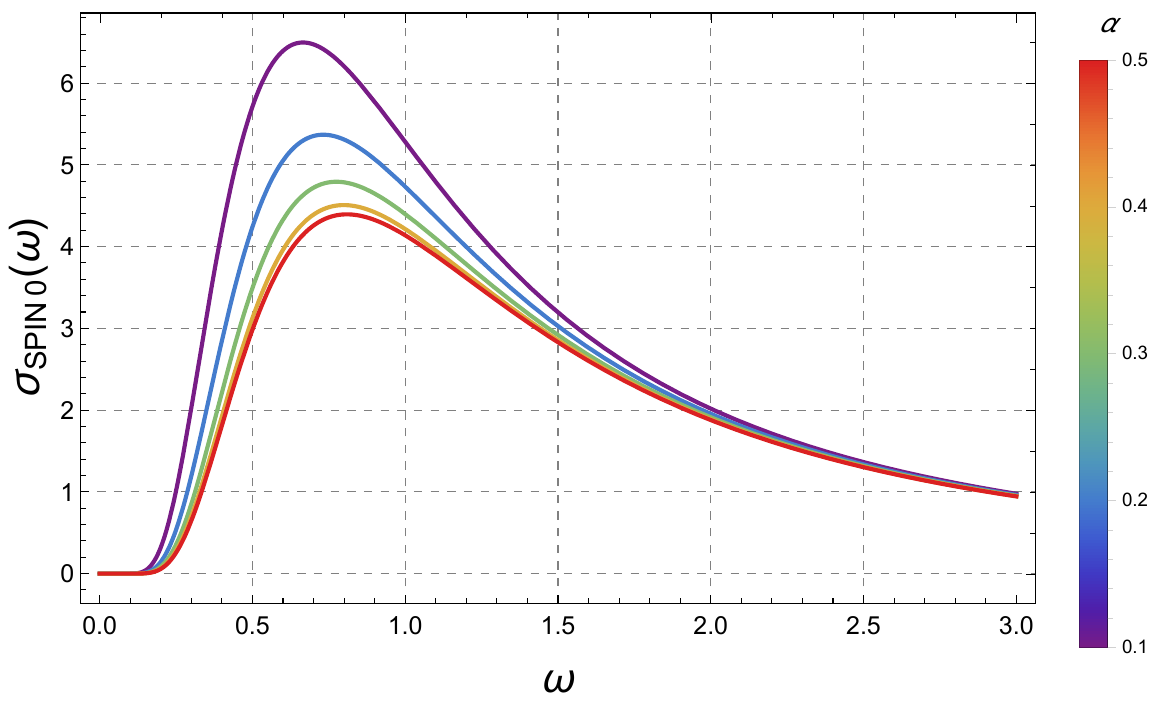}

    \vspace{0.15cm}
    \parbox{0.95\textwidth}{\centering\footnotesize
    (d) $\sigma_0(\omega)$: same parameters as in (b)}
\end{minipage}

\vspace{-0.2cm}

\caption{Behavior of the greybody lower bound and absorption cross section for spin $0$. Top row: transmission $T_0(\omega)$ \eqref{eq:T0}; bottom row: absorption cross section $\sigma_0(\omega)$ \eqref{eq:sigma}. Both $Q$ and $\alpha$ suppress transmission and absorption by raising the effective barrier.}
\label{TA0}
\end{figure*}

\subsection{Spin-1}

\begin{figure*}[ht!]
\centering
\begin{minipage}[t]{0.48\textwidth}
    \centering
    \includegraphics[height=4.8cm]{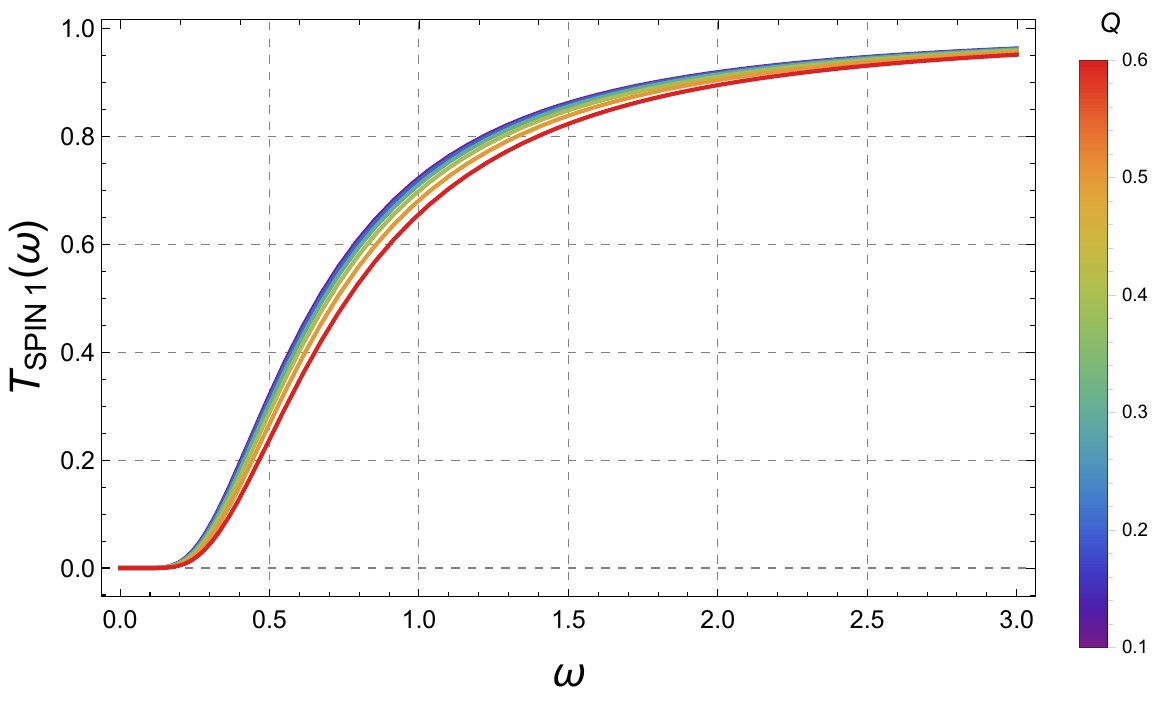}

    \vspace{0.15cm}
    \parbox{0.95\textwidth}{\centering\footnotesize
    (a) $T_1(\omega)$: varying $Q\in\{0.1,\ldots,0.6\}$; 
    $\alpha=0.01,\,l=1,\,M=1$}
\end{minipage}
\hfill
\begin{minipage}[t]{0.48\textwidth}
    \centering
    \includegraphics[height=4.8cm]{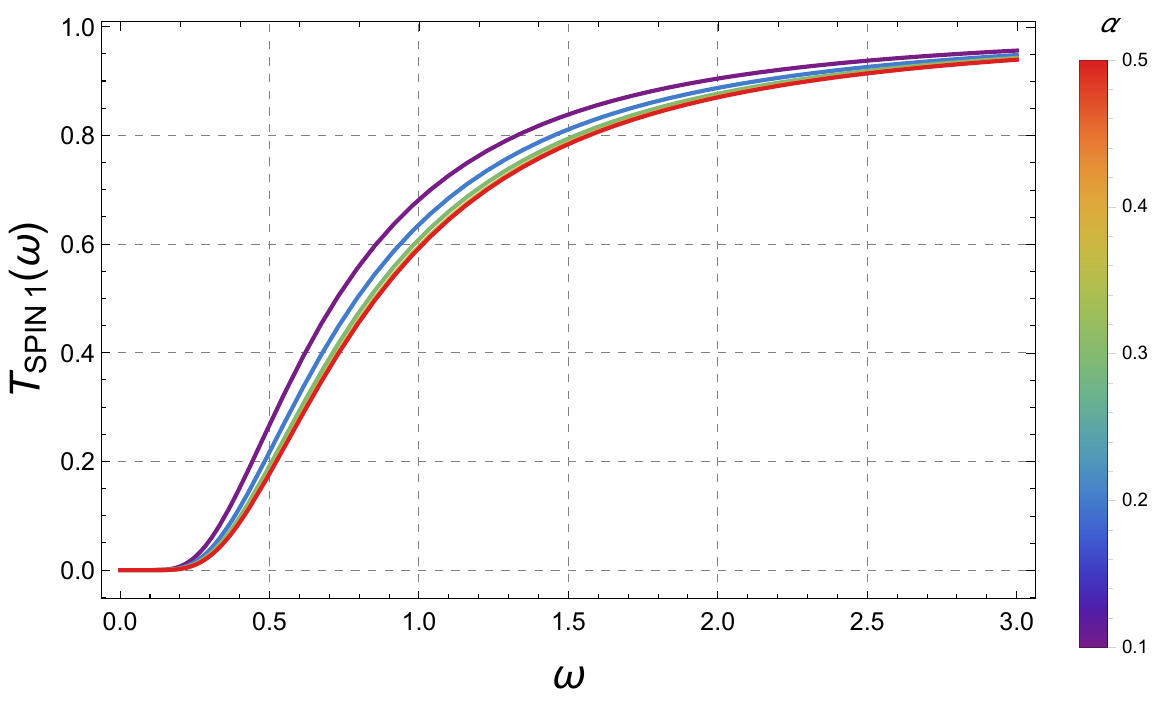}

    \vspace{0.15cm}
    \parbox{0.95\textwidth}{\centering\footnotesize
    (b) $T_1(\omega)$: varying $\alpha\in\{0.1,\ldots,0.5\}$; 
    $Q=0.1,\,l=1,\,M=1$}
\end{minipage}

\vspace{0.45cm}

\begin{minipage}[t]{0.48\textwidth}
    \centering
    \includegraphics[height=4.8cm]{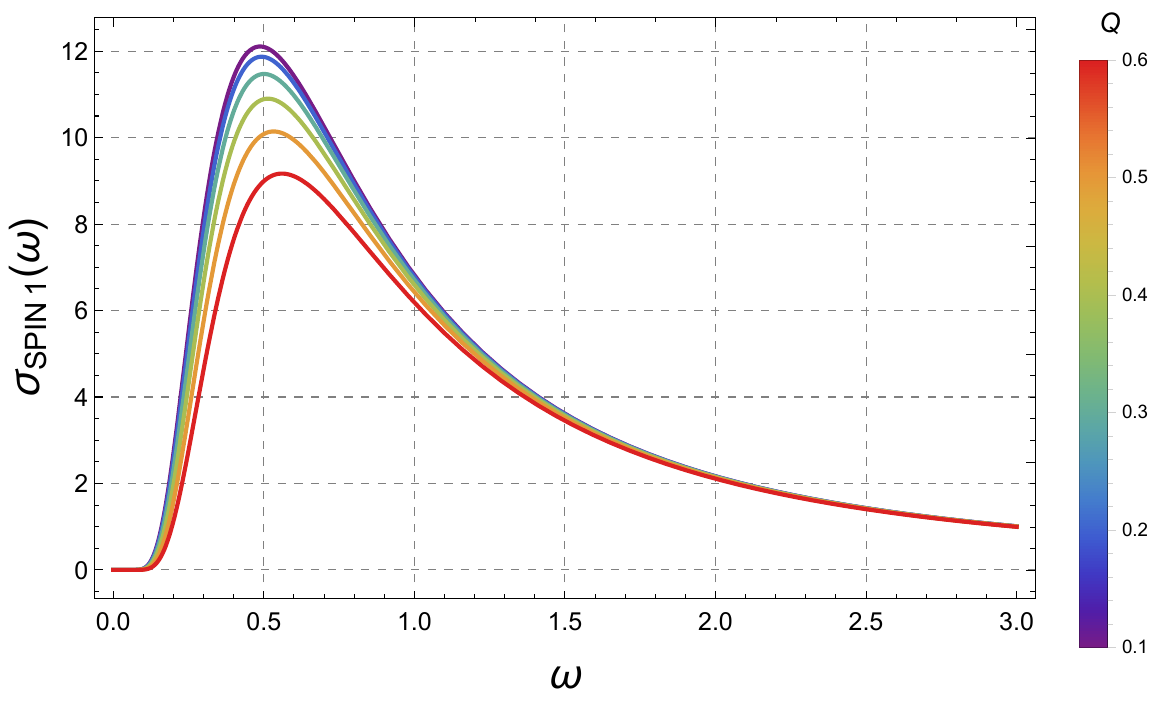}

    \vspace{0.15cm}
    \parbox{0.95\textwidth}{\centering\footnotesize
    (c) $\sigma_1(\omega)$: same parameters as in (a)}
\end{minipage}
\hfill
\begin{minipage}[t]{0.48\textwidth}
    \centering
    \includegraphics[height=4.8cm]{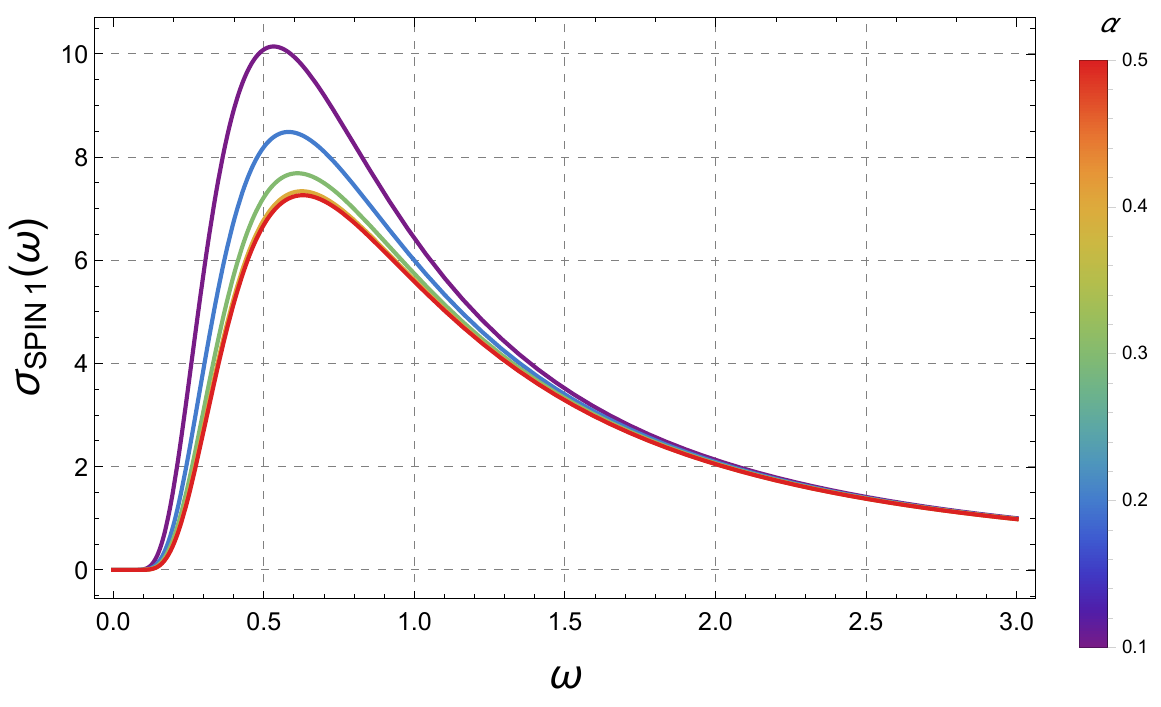}

    \vspace{0.15cm}
    \parbox{0.95\textwidth}{\centering\footnotesize
    (d) $\sigma_1(\omega)$: same parameters as in (b)}
\end{minipage}

\vspace{-0.2cm}

\caption{Behavior of the greybody lower bound and absorption cross section for spin $1$. Top row: transmission $T_1(\omega)$ \eqref{eq:T1}; bottom row: absorption cross section $\sigma_1(\omega)$ \eqref{eq:sigma}.}
\label{TA1}
\end{figure*}

For the spin 1 case, we obtain the following lower bound for the greybody factor written in terms of $r_h$
\begin{align}
    T_{\rm spin\,1}=\mathrm{sech}^2\!\left[\frac{1}{2\omega}\frac{l(l+1)}{r_h}\right].
    \label{eq:T1}
\end{align}
Upon analyzing the behavior of the greybody lower bound for spin 1 in Fig.~\ref{TA1}, we note that the vector case displays the same overall trend, namely a monotonic increase of the lower bound $T_{\rm spin\,1}(\omega)$ with $\omega$ and a reduction of the transmission when either $Q$ or $\alpha$ is increased. However, the dependence is simpler than in the scalar sector, since the vector transmission is controlled essentially by the inverse horizon radius (see Eq.~\eqref{eq:T1}). This is physically reasonable: once the horizon shrinks, the effective barrier becomes wider in tortoise coordinates, and the transmission probability decreases. Even though the qualitative behavior is the same as in the scalar case, the vector greybody lower bound is systematically larger, which means that spin-1 modes are less suppressed by the background than scalar ones in the frequency range explored.

\subsection{Spin-1/2}

\begin{figure*}[ht!]
\centering
\begin{minipage}[t]{0.48\textwidth}
    \centering
    \includegraphics[height=4.8cm]{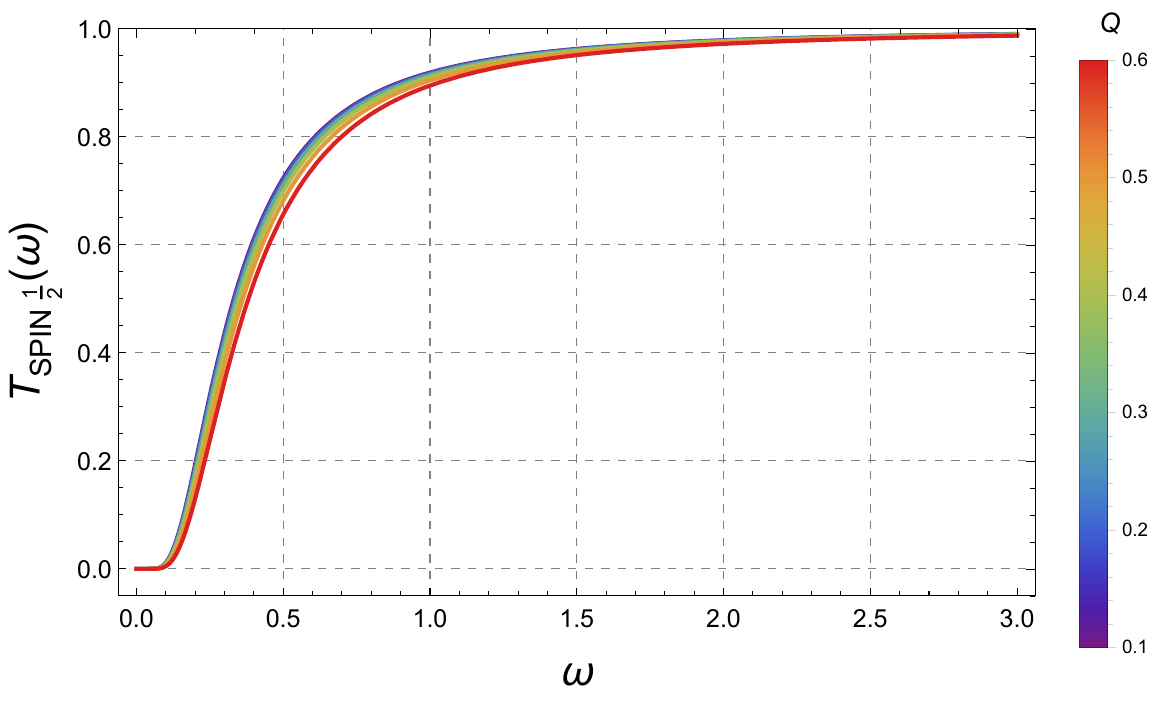}
    \vspace{0.15cm}
    \footnotesize
    (a) $T_{1/2}(\omega)$: varying $Q\in\{0.1,\ldots,0.6\}$; 
    $\alpha=0.01,\,\kappa=1,\,M=1$
\end{minipage}
\hfill
\begin{minipage}[t]{0.48\textwidth}
    \centering
    \includegraphics[height=4.8cm]{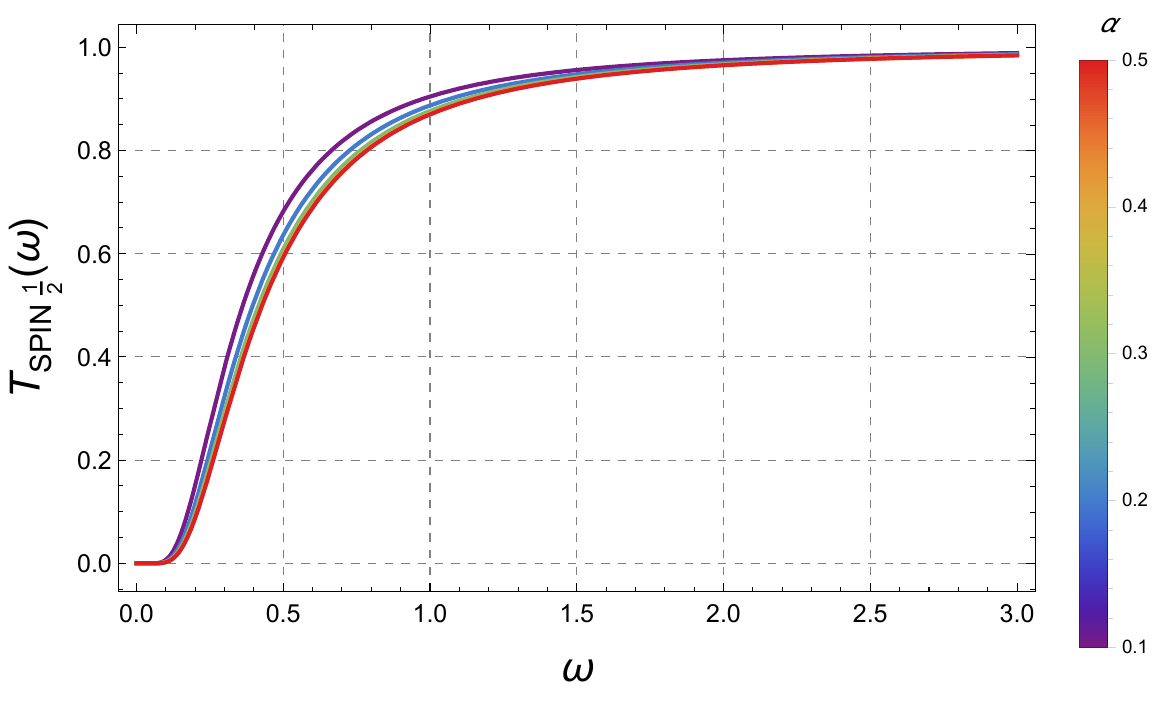}
    \vspace{0.15cm}
    \footnotesize
    (b) $T_{1/2}(\omega)$: varying $\alpha\in\{0.1,\ldots,0.5\}$; 
    $Q=0.1,\,\kappa=1,\,M=1$
\end{minipage}
\vspace{0.45cm}

\begin{minipage}[t]{0.48\textwidth}
    \centering
    \includegraphics[height=4.8cm]{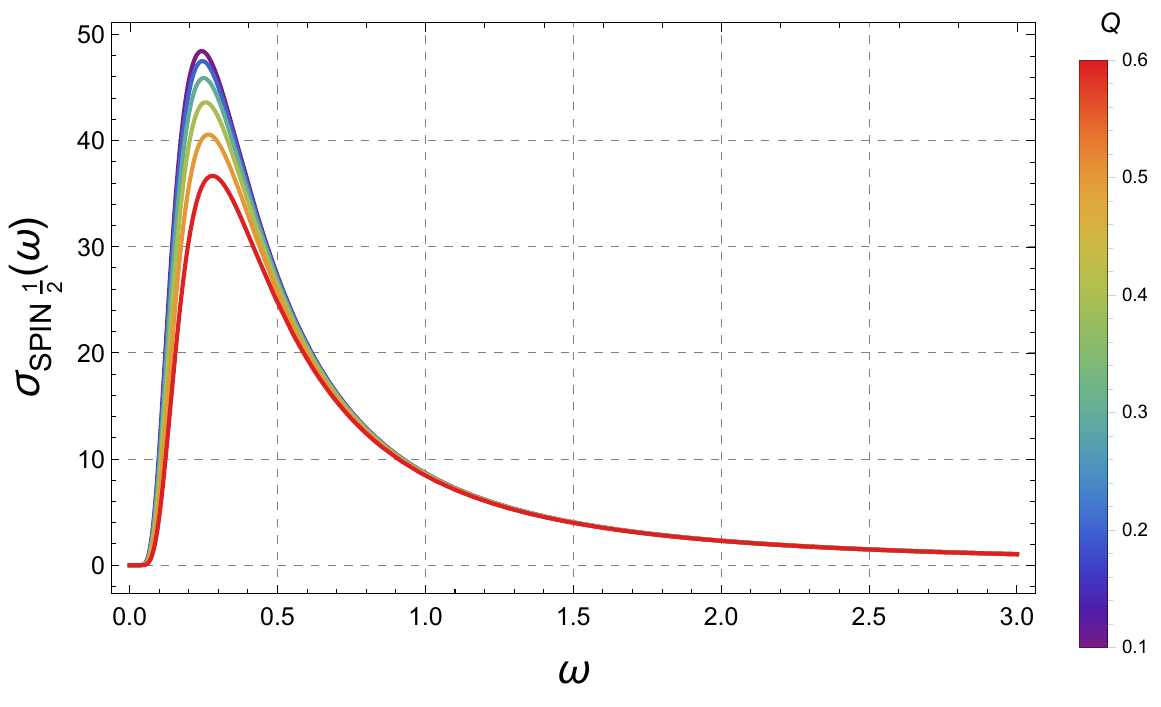}
    \vspace{0.15cm}
    \footnotesize
    (c) $\sigma_{1/2}(\omega)$: same parameters as in (a)
\end{minipage}
\hfill
\begin{minipage}[t]{0.48\textwidth}
    \centering
    \includegraphics[height=4.8cm]{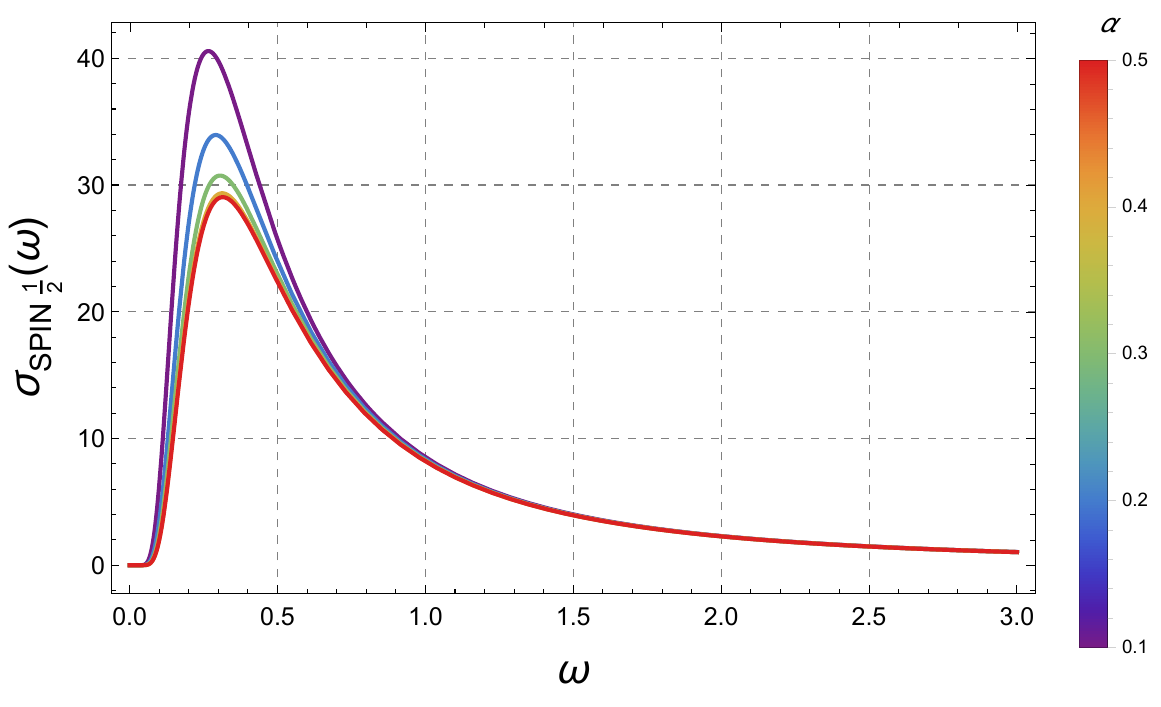}

    \vspace{0.15cm}
    \footnotesize
    (d) $\sigma_{1/2}(\omega)$: same parameters as in (b)
\end{minipage}

\vspace{-0.2cm}

\caption{Behavior of the greybody lower bound and absorption cross section for spin $1/2$. Top row: transmission $T_{1/2}(\omega)$ \eqref{eq:T12}; bottom row: absorption cross section $\sigma_{1/2}(\omega)$ \eqref{eq:sigma}. The Dirac angular quantum number $\kappa=1$ is used throughout.}
\label{TA12}
\end{figure*}

The Boonserm-Visser lower bound for the spin 1/2 greybody factor in terms of $r_h$ reads
\begin{align}
    T_{\rm spin\,1/2}=\mathrm{sech}^2\!\left[\frac{1}{2\omega}\frac{\kappa^2}{r_h}\right],
    \label{eq:T12}
\end{align}
where $\kappa=\pm(l+\tfrac{1}{2}+\tfrac{1}{2})=\pm(l+1)$ is the Dirac angular quantum number introduced in Section~\ref{s3}. In Fig.~\ref{TA12}, the greybody lower bound for the fermionic field $T_{\rm spin\,1/2}(\omega)$ is depicted and again shows an increasing function of the frequency, but it remains above the scalar and vector curves for the same choice of parameters. This indicates that spin-$1/2$ particles cross the effective barrier more efficiently. The corresponding barrier integral $I_{1/2}=\kappa^2/r_h$ is smaller than $I_1=l(l+1)/r_h$ for $\kappa^2 < l(l+1)$ (e.g., $\kappa=1$, $l=1$: $I_{1/2}=1/r_h < 2/r_h = I_1$), which naturally explains the larger transmission. Therefore, within the representative parameter range and angular choices explored here, the comparison among the three channels reveals the ordering,
\begin{equation}
T_{\rm spin\,1/2}(\omega) > T_{\rm spin\,1}(\omega) > T_{\rm spin\,0}(\omega),
\label{eq:ordering}
\end{equation}
for the representative set of parameters used in the plots. This hierarchy is not asserted as a universal theorem for all possible angular numbers, chiral sectors, and parameter values; rather, it summarizes the behavior of the configurations studied here. As the frequency increases, all three curves tend to unity, and the differences between the spins become less pronounced, which is exactly what one expects once the potential barrier becomes negligible compared with the wave energy.

The absorption cross section presents a richer structure. In all spin sectors, $\sigma(\omega)$ vanishes in the very low-frequency region, rises to a maximum at intermediate frequencies, and then decreases again for large $\omega$. This non-monotonic behavior has a simple physical origin. At small $\omega$, the transmission probability is exponentially suppressed, so the black hole absorbs only a tiny fraction of the incoming radiation. At high frequencies, however, although the exact greybody factor, and correspondingly the bound in the high-frequency regime, tends to one, the overall prefactor $1/\omega^{2}$ in Eq.~\eqref{eq:sigma} forces the absorption cross section to decay. The competition between these two effects produces a single broad peak in the intermediate regime.

The comparison between the three absorption channels is particularly illuminating. For the same background parameters and angular choice used above, the fermionic cross section is the largest, the vector one lies in the middle, and the scalar one is the smallest. Thus, the ordering found for the greybody lower bounds is preserved at the level of absorption:
\begin{equation}
\sigma_{\rm spin\,1/2}(\omega) > \sigma_{\rm spin\,1}(\omega) > \sigma_{\rm spin\,0}(\omega).
\end{equation}
This result shows that the field's spin plays a decisive role in the interaction between the Hawking quanta and the effective geometry outside the horizon. Fermions are less hindered by the barrier and therefore dominate the absorption spectrum over the parameter range considered, whereas scalar modes are most strongly suppressed.

Overall, these results support a coherent physical interpretation: the electric charge $Q$ and the dark matter parameter $\alpha$ act together to reduce the horizon radius and strengthen the geometry's filtering effect. As a consequence, the escape of Hawking radiation becomes less efficient, and the absorption of external waves is weakened. At the same time, the field spin controls the strength of each channel's response to this filtering. Scalar modes are the most sensitive to the combined geometric and matter effects, vector modes are moderately affected, and fermionic modes remain the least suppressed. These results suggest that both nonlinear electrodynamic corrections and the surrounding dark matter distribution can leave diagnostic imprints on the spectral profile of black hole radiation.

\subsection{Effect of the Euler-Heisenberg parameter}

\begin{figure*}[ht!]
\begin{center}
\includegraphics[scale=0.55]{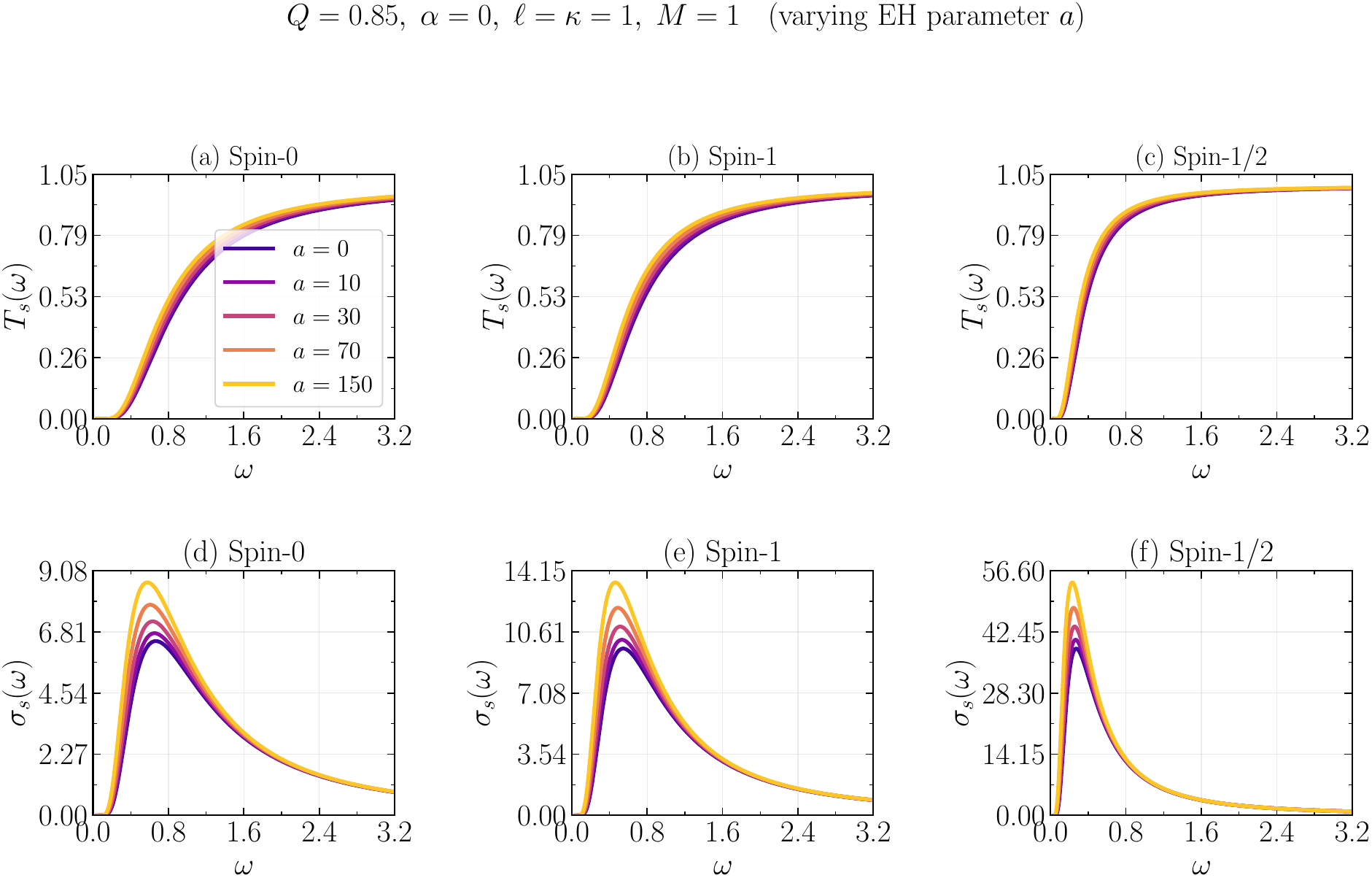}
\end{center}
\vspace{-0.4cm}
\caption{Greybody lower bound $T_s(\omega)$ (top row: a, b, c) and absorption cross section $\sigma_s(\omega)$ (bottom row: d, e, f) for varying EH parameter $a\in\{0,10,30,70,150\}$. Parameters: $M=1$, $Q=0.85$, $\alpha=0$, $l=\kappa=1$. Columns from left to right: spin-0, spin-1, spin-1/2. Larger $a$ pushes $r_h$ from 1.53 to 1.80 and therefore increases $\Gamma_s$ --- opposite to the effect of $Q$ and $\alpha$.}
\label{fig:EH_param}
\end{figure*}

The analyses in Subsections A through C hold the EH nonlinearity parameter $a$ fixed at a small reference value. Figure~\ref{fig:EH_param} isolates its role by setting $Q=0.85$ (so that $aQ^4$ is non-negligible) and $\alpha=0$ (no PFDM), while varying $a\in\{0,10,30,70,150\}$. This range is deliberately large and purely illustrative; for astrophysically realistic values ($a\sim 10^{-30}$ in Planck units), the EH correction is negligible and all curves collapse onto the $a=0$ line.

As Fig.~\ref{fig:EH_param} shows, increasing $a$ systematically \emph{increases} the greybody lower bound and absorption cross section for all three spin sectors. This behavior is opposite to the suppression produced by $Q$ and $\alpha$, and has a direct geometric explanation: the EH correction $-aQ^4/(20r^6)$ in Eq.~\eqref{eq:metric} is negative, which lowers $g(r)$ near the singularity and shifts the horizon outward. As $r_h$ grows, the integrated barrier $I_s \propto 1/r_h$ decreases and the Boonserm-Visser bound $\Gamma_s \geq \mathrm{sech}^2(I_s/2\omega)$ yields larger transmission. Numerically, the horizon grows from $r_h \approx 1.53$ at $a=0$ to $r_h \approx 1.80$ at $a=150$ (a variation of about 17\%), which produces the visible separation of the curves in Fig.~\ref{fig:EH_param}.

These results reveal a qualitative distinction between the charge effect (larger $Q$ $\Rightarrow$ smaller $r_h$ $\Rightarrow$ stronger barrier $\Rightarrow$ smaller $\Gamma_s$) and the EH vacuum-polarisation effect (larger $a$ $\Rightarrow$ larger $r_h$ $\Rightarrow$ weaker barrier $\Rightarrow$ larger $\Gamma_s$). While the EH correction is negligible in practice, identifying its qualitative direction is important for understanding how quantum electrodynamic effects and dark-matter environments compete in shaping the transmission spectrum.

\subsection{Spin comparison and model limits}

\begin{figure*}[ht!]
\begin{center}
\includegraphics[scale=0.55]{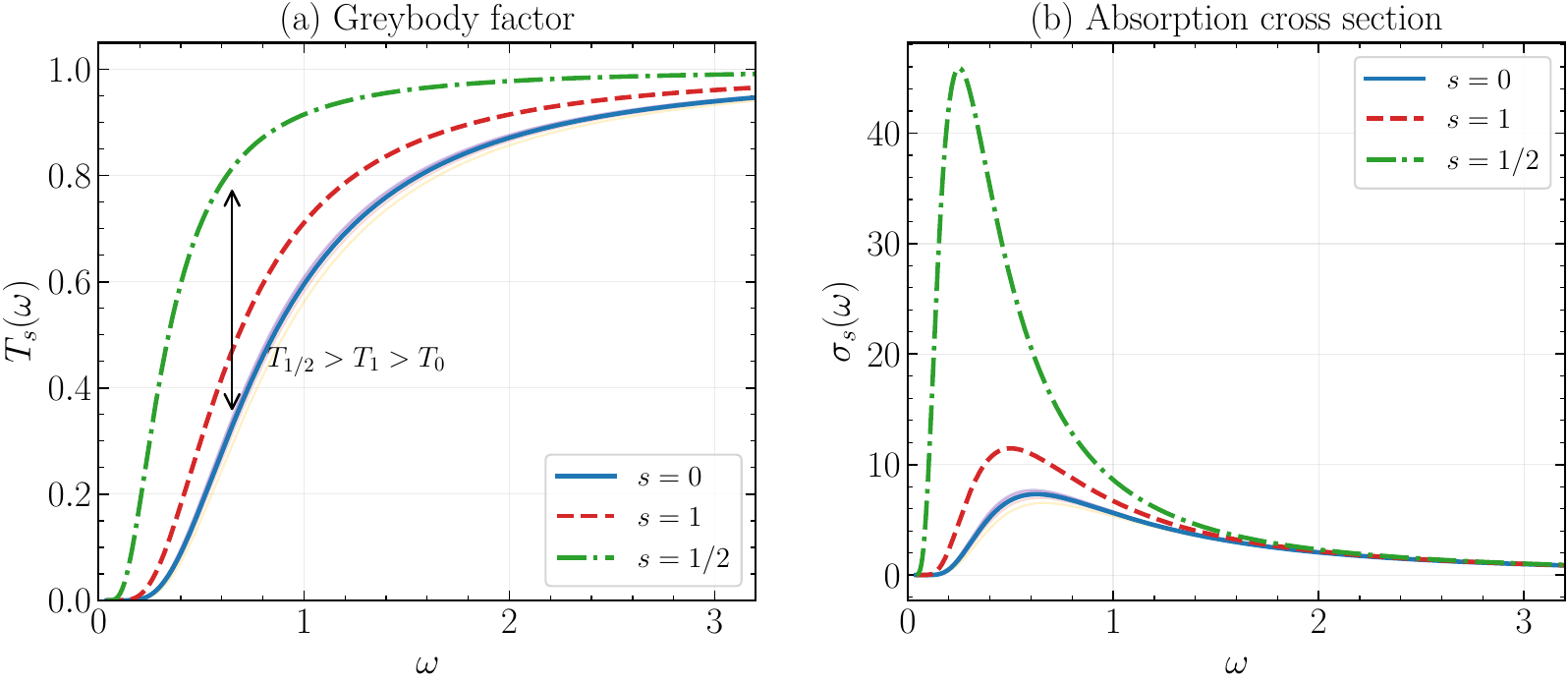}
\end{center}
\vspace{-0.4cm}
\caption{Greybody lower bound $T_s(\omega)$ (a) and absorption cross section $\sigma_s(\omega)$ (b) for all three spin sectors: $s=0$ (solid), $s=1$ (dashed), $s=1/2$ (dash-dotted). Parameters: $M=1$, $Q=0.3$, $a=1.0$, $\alpha=0.1$, $l=\kappa=1$. Faint background curves show the $Q$-dependence of $T_0$ for $Q\in\{0.1,0.2,0.4,0.5\}$. The ordering $T_{1/2}>T_1>T_0$ \eqref{eq:ordering} holds throughout the plotted parameter range.}
\label{fig:spin_comp}
\end{figure*}

Figure~\ref{fig:spin_comp} displays the greybody lower bound and absorption cross section for all three spin sectors simultaneously, for reference parameters $M=1$, $Q=0.3$, $a=1.0$, $\alpha=0.1$, $l=\kappa=1$. The faint background curves in both panels show the dependence of the scalar ($s=0$) channel on $Q\in\{0.1,0.2,0.4,0.5\}$, providing a visual sense of the parameter sensitivity. The main curves confirm the ordering \eqref{eq:ordering} across the explored frequency range. The ordering follows directly from the integrated barriers: $I_{1/2}=\kappa^2/r_h=1/r_h$, $I_1=l(l+1)/r_h=2/r_h$, and $I_0 > I_1$ because the scalar integral contains the additional curvature-coupling contribution from $f'(r)/r$. As $\omega\to\infty$, all three curves converge to unity and the spin differences vanish, consistent with the geometric-optics limit. The same ordering is preserved for $\sigma_s$.

\begin{figure*}[ht!]
\begin{center}
\includegraphics[scale=0.55]{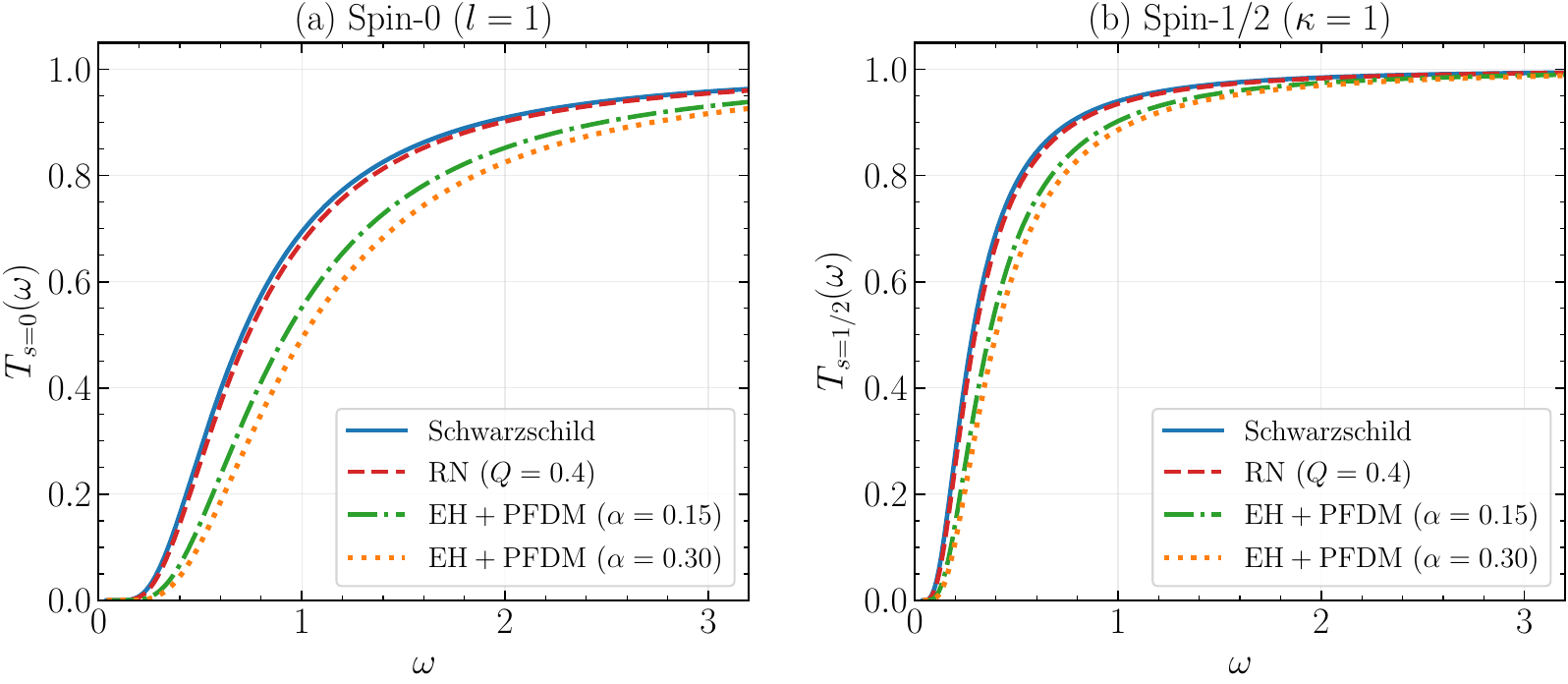}
\end{center}
\vspace{-0.4cm}
\caption{Model comparison of the greybody lower bound for spin-0 (a) and spin-1/2 (b), with $M=1$, $Q=0.4$, $a=0.01$, $l=\kappa=1$. Shown are: Schwarzschild ($r_h=2$), RN ($r_h\approx1.92$), EH+PFDM with $\alpha=0.15$ ($r_h\approx1.55$), and EH+PFDM with $\alpha=0.30$ ($r_h\approx1.42$). The hierarchy $T_{\rm Schw}>T_{\rm RN}>T_{\rm EH+PFDM}$ confirms that the PFDM logarithmic term is the dominant modifier of the greybody spectrum.}
\label{fig:model_comp}
\end{figure*}

Figure~\ref{fig:model_comp} places the EH+PFDM results in the context of their known limits by comparing four models: Schwarzschild ($Q=\alpha=a=0$), Reissner-Nordstr\"{o}m (RN, $\alpha=a=0$, $Q=0.4$), and EH+PFDM with $\alpha=0.15$ and $\alpha=0.30$, all with $M=1$, $Q=0.4$, $a=0.01$, $l=\kappa=1$. The hierarchy
\begin{align}
T_s^{\rm Schw} &> T_s^{\rm RN} > T_s^{\rm EH+PFDM}(\alpha=0.15)\notag\\& > T_s^{\rm EH+PFDM}(\alpha=0.30)
\end{align}
is clearly visible for both spin-0 and spin-1/2. The Schwarzschild background ($r_h=2M=2$) transmits the most; adding charge ($r_h\approx1.92$) and PFDM ($r_h\approx1.55$ for $\alpha=0.15$ and $r_h\approx1.42$ for $\alpha=0.30$) progressively suppresses transmission. These results confirm that the Schwarzschild and RN limits are correctly recovered, and that the PFDM logarithmic term drives the principal model-dependent departure from the RN baseline, while the EH correction ($a=0.01$) is negligible for the parameter values shown.

\section{Energy emission}

This section studies the Hawking energy emission spectrum of massless fields propagating in a charged EH black hole background surrounded by PFDM. In this context, the energy-emission spectrum is important, as it indicates how much energy is carried away by the Hawking flux and is an essential observable for understanding the evaporation process in a complete way. Its behavior can be determined by the Hawking temperature and the greybody factors, which encode the influence of spacetime geometry, the effective potential barrier, and the properties of the emitted field, as we discussed in the previous section. In this sense, the analysis of energy emission provides a powerful bridge among thermodynamics, quantum field theory in curved spacetime, and the observable properties of black holes.

In technical terms, we write the spectral energy emission rate as follows
\begin{align}
\frac{d^2E}{d\omega\,dt}
&=\frac{2\pi^2\omega^3}{e^{\omega/T_H}-1}\,
\sigma_m(\omega),
\end{align}
for bosonic fields and
\begin{align}
\frac{d^2E}{d\omega\,dt}
&=\frac{2\pi^2\omega^3}{e^{\omega/T_H}+1}\,
\sigma_m(\omega),
\end{align}
for fermionic fields. Here, $T_H$ is the Hawking temperature and $\sigma_m(\omega)$ is the partial absorption cross section associated with the field mode. Let us calculate the Hawking temperature explicitly. The event horizon \(r_h\) is defined by the largest positive solution of $f(r_h)=0$. All emission quantities in this analysis are evaluated at this outer horizon. Then, the Hawking temperature follows from the surface gravity, namely
\begin{equation}
    T_H=\frac{f'(r_h)}{4\pi},
\end{equation}
which gives
\begin{align}
T_H=&\frac{3aQ^{4}+10r_h^{4}\left(-2Q^{2}+2Mr_h+\alpha r_h\right)
}{40\pi r_h^{7}}\nonumber\\&-\frac{10\alpha r_h^{5}\ln\!\left(r_h/\alpha\right)}{40\pi r_h^{7}} .
    \label{eq:temperature-EH-PFDM}
\end{align}
This expression already contains much of the relevant physics. The Maxwell charge term tends to reduce the temperature, pushing the configuration toward a colder, near-extremal regime.  The EH contribution appears through the higher-order term \(aQ^{4}\), so its influence is important only when the electromagnetic sector is sufficiently strong or when the horizon is small enough for inverse powers of \(r_h\) to be amplified. The PFDM parameter contributes through both a linear term in \(\alpha\) and a logarithmic term, so its effect depends on the ratio \(r_h/\alpha\), not solely on the absolute magnitude of \(\alpha\).

With this expression for the Hawking temperature and the expressions for the greybody lower bound and absorption cross-section, we can plot and analyze the behavior of the energy emission rate for spin 0, 1, and 1/2 particles.
\begin{figure*}[ht!]
\centering

\begin{minipage}[t]{0.48\textwidth}
    \centering
    \includegraphics[height=4.8cm]{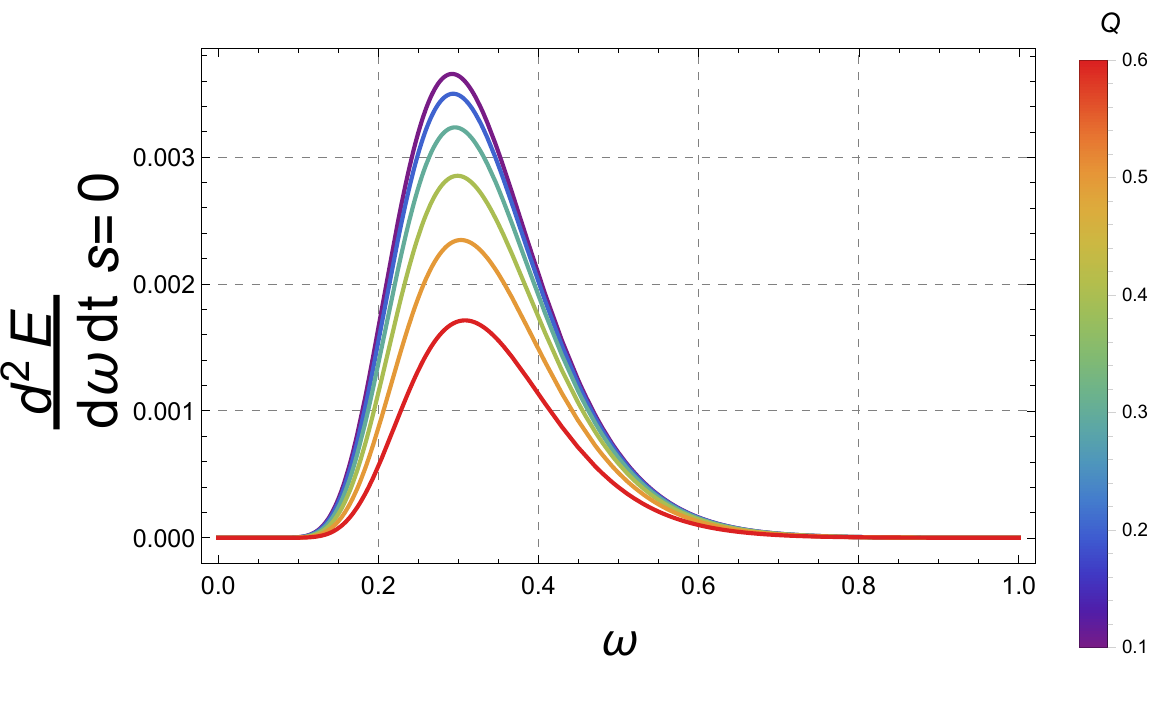}

    \vspace{0.15cm}
    \footnotesize
    (a) $\frac{d^2E}{d\omega dt}_{spin 0}(\omega)$: varying $Q\in\{0.1,\ldots,0.6\}$; 
    $\alpha=0.01,\,l=1,\,M=1$
\end{minipage}
\hfill
\begin{minipage}[t]{0.48\textwidth}
    \centering
    \includegraphics[height=4.8cm]{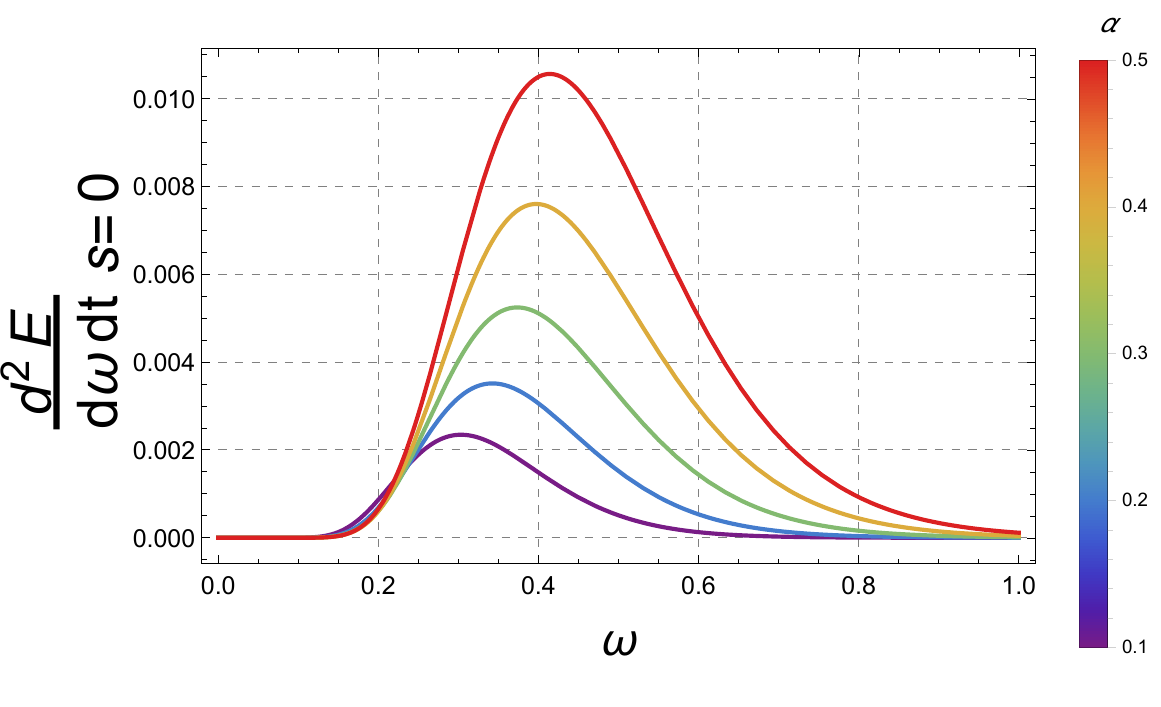}

    \vspace{0.15cm}
    \footnotesize
    (b) $\frac{d^2E}{d\omega dt}_{spin 0}(\omega)$: varying $\alpha\in\{0.1,\ldots,0.5\}$; 
    $Q=0.1,\,l=1,\,M=1$
\end{minipage}

\vspace{-0.2cm}

\caption{Behavior of the energy emission rate for spin $0$. The spin-0 angular quantum number $l=1$ is used throughout.}
\label{E0}
\end{figure*}

\begin{figure*}[ht!]
\centering

\begin{minipage}[t]{0.48\textwidth}
    \centering
    \includegraphics[height=4.8cm]{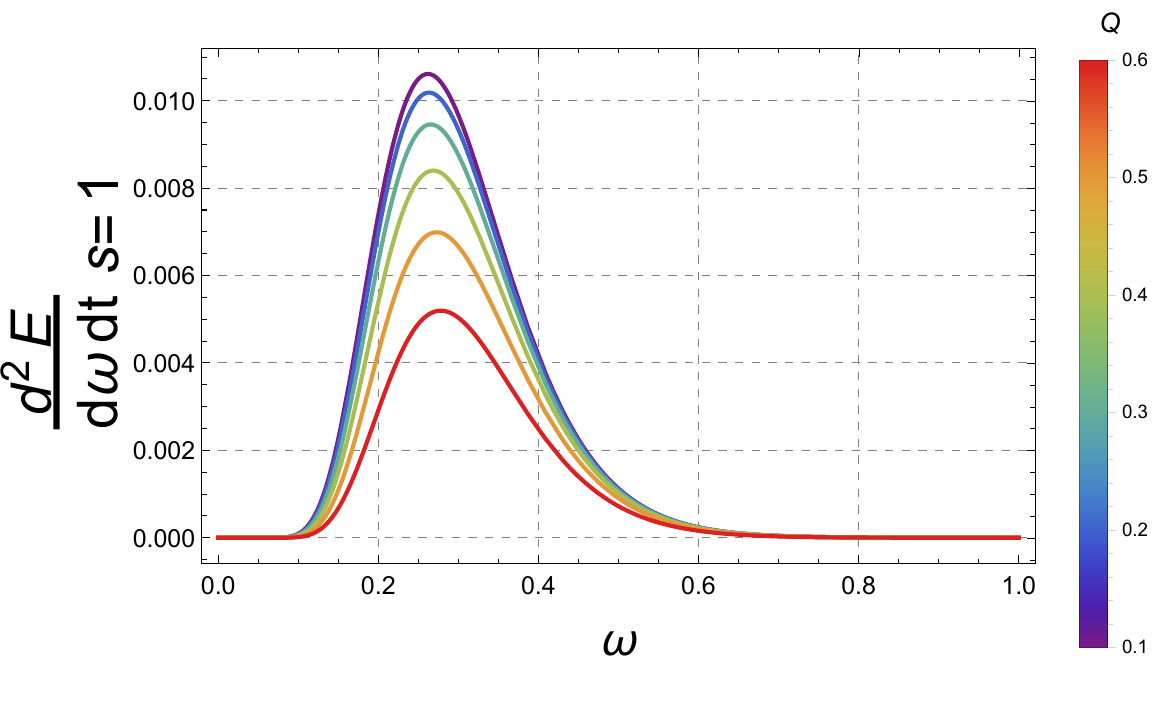}

    \vspace{0.15cm}
    \footnotesize
    (a) $\frac{d^2E}{d\omega dt}_{spin 1}(\omega)$: varying $Q\in\{0.1,\ldots,0.6\}$; 
    $\alpha=0.01,\,l=1,\,M=1$
\end{minipage}
\hfill
\begin{minipage}[t]{0.48\textwidth}
    \centering
    \includegraphics[height=4.8cm]{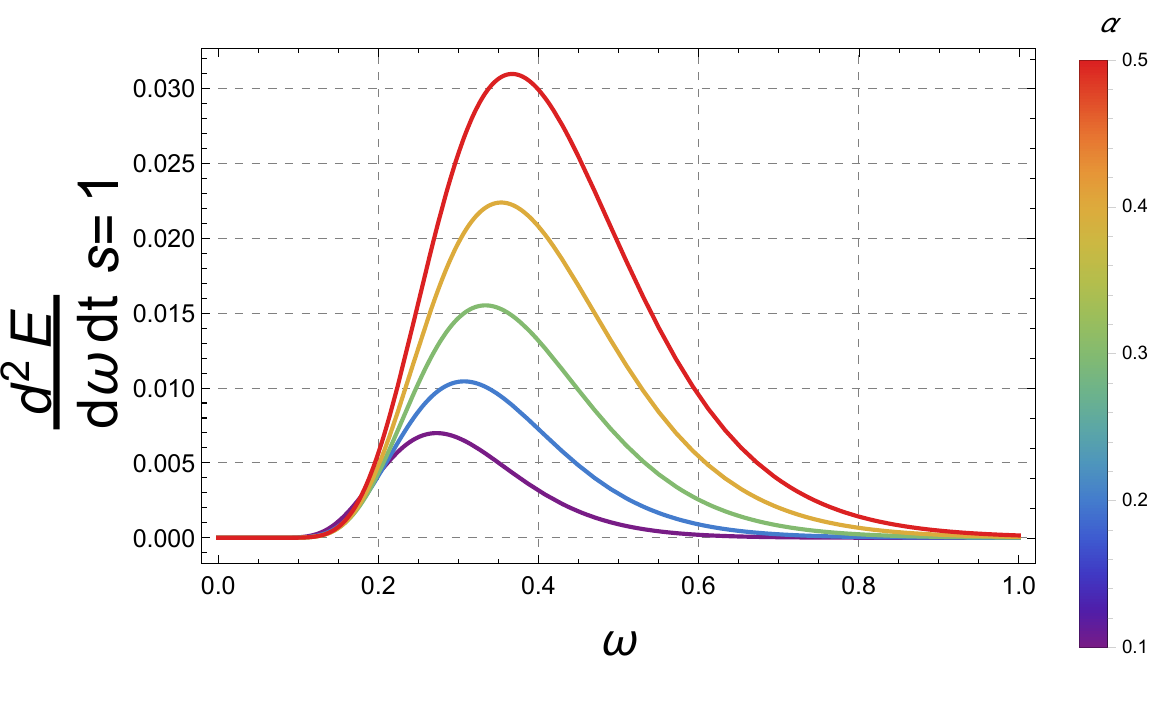}

    \vspace{0.15cm}
    \footnotesize
    (b) $\frac{d^2E}{d\omega dt}_{spin 1}(\omega)$: varying $\alpha\in\{0.1,\ldots,0.5\}$; 
    $Q=0.1,\,l=1,\,M=1$
\end{minipage}

\vspace{-0.2cm}

\caption{Behavior of the energy emission rate for spin $1$. The spin-1 angular quantum number $l=1$ is used throughout.}
\label{E1}
\end{figure*}

\begin{figure*}[ht!]
\centering

\begin{minipage}[t]{0.48\textwidth}
    \centering
    \includegraphics[height=4.8cm]{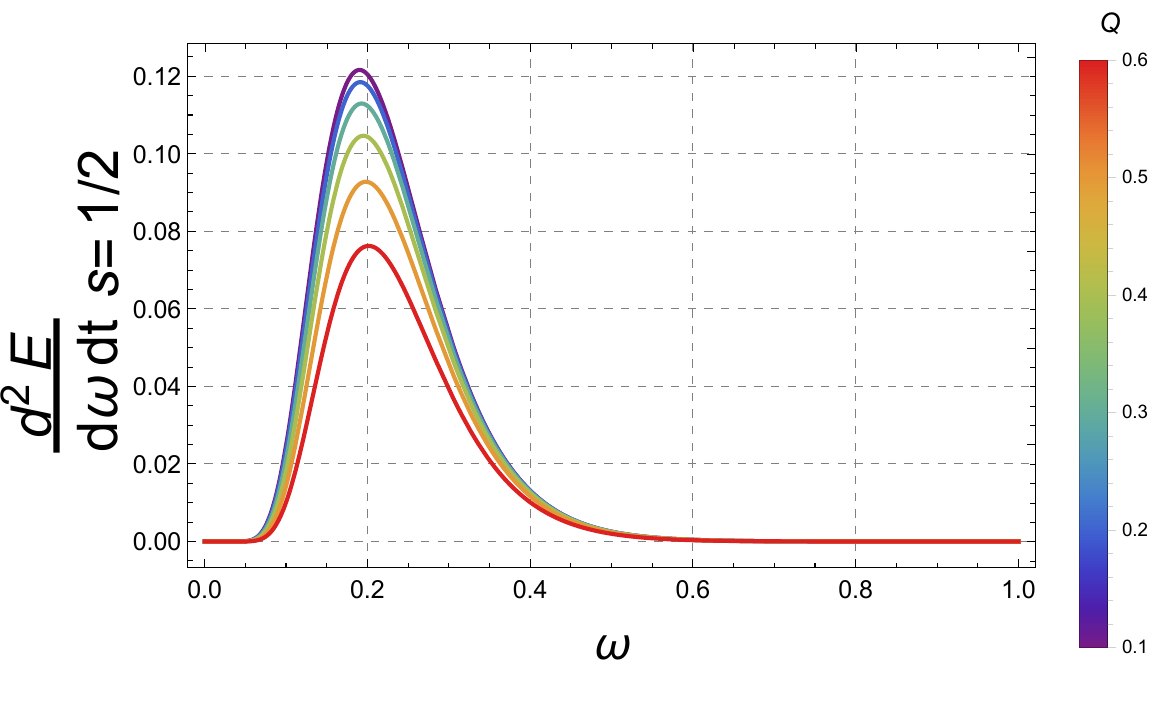}

    \vspace{0.15cm}
    \footnotesize
    (a) $\frac{d^2E}{d\omega dt}_{spin 1/2}(\omega)$: varying $Q\in\{0.1,\ldots,0.6\}$; 
    $\alpha=0.01,\,\kappa=1,\,M=1$
\end{minipage}
\hfill
\begin{minipage}[t]{0.48\textwidth}
    \centering
    \includegraphics[height=4.8cm]{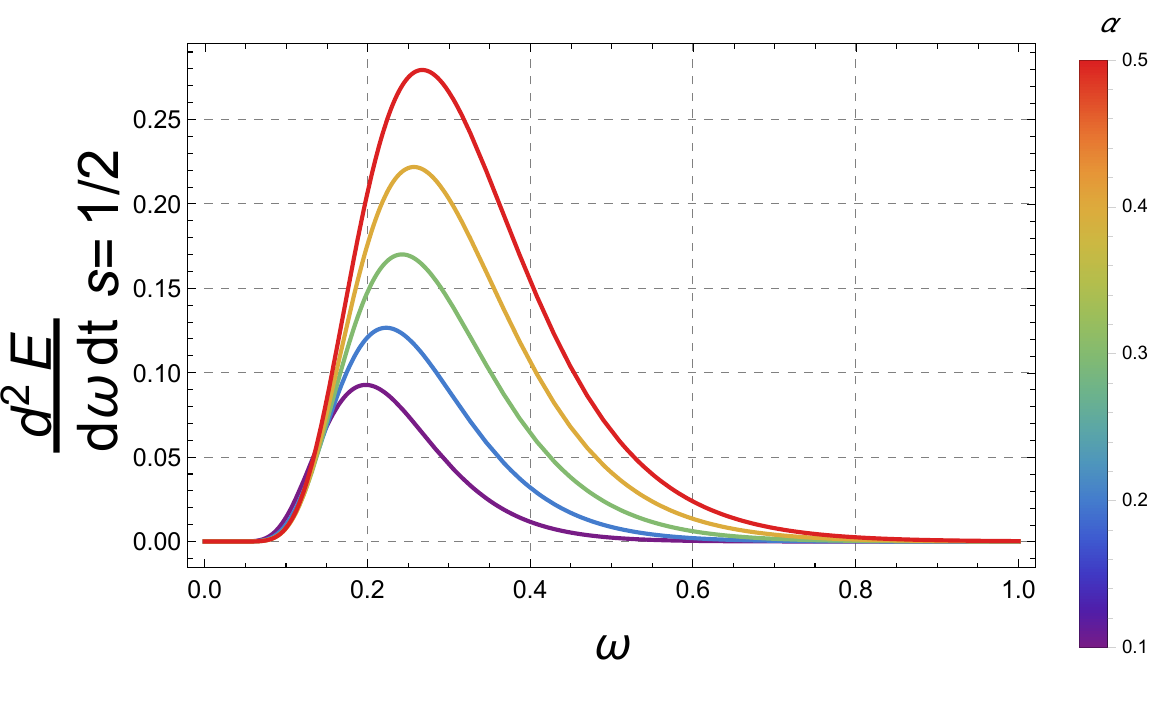}

    \vspace{0.15cm}
    \footnotesize
    (b) $\frac{d^2E}{d\omega dt}_{spin 1/2}(\omega)$: varying $\alpha\in\{0.1,\ldots,0.5\}$; 
    $Q=0.1,\,\kappa=1,\,M=1$
\end{minipage}

\vspace{-0.2cm}

\caption{Behavior of the energy emission rate for spin $1/2$. The Dirac angular quantum number $\kappa=1$ is used throughout.}
\label{E12}
\end{figure*}

We plot the behavior of the energy emission rate for spin $0$, $1$, and $1/2$ in Figs. (\ref{E0})-(\ref{E12}). In the numerical range considered in these plots, we see that increasing \(\alpha\) reduces the outer horizon radius and increases the Hawking temperature. This is an important point because the PFDM term is logarithmic. Its effect cannot be inferred from a single power-law correction. Instead, the sign and strength of the contribution depend on \(\ln(r_h/\alpha)\), and then on the relative size of the horizon compared with the dark-matter scale. The plots show that increasing \(\alpha\) enhances the emission rate for spin \(0\), spin \(1\), and spin \(1/2\).  The peaks become higher and move to larger frequencies.  This is exactly what one expects when the black hole becomes hotter, i.e., when \(T_H\) increases, shifting the thermal spectrum to the right and raising the emitted flux.  Although the PFDM term also affects the greybody lower bound through the horizon radius, the thermal enhancement dominates over the parameter range used in the notebook.

The scalar channel again carries the most detailed imprint of the geometry. Note that the PFDM correction appears as
\begin{equation}
\frac{\alpha}{4r_h^{2}}-\frac{\alpha}{2r_h^{2}}
\ln\!\left(\frac{r_h}{\alpha}\right).
\end{equation}
This term may either increase or decrease the scalar barrier depending on the value of \(r_h/\alpha\). Thus, PFDM does not merely amplify the emission in a universal way. Instead, it can change both the temperature and the transmission probability. In the specific numerical domain of the notebook, the net result is an enhancement of the radiative spectrum.

\begin{figure*}[ht!]
\centering

\begin{minipage}[t]{0.48\textwidth}
    \centering
    \includegraphics[height=4.8cm]{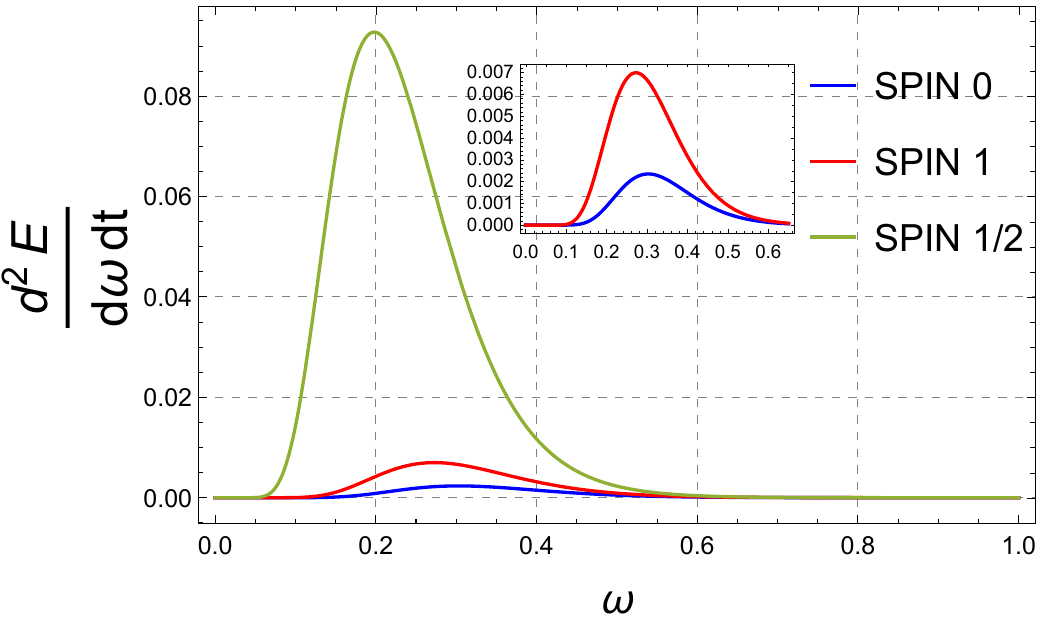}

    \vspace{0.15cm}
    \footnotesize
    (a) $\frac{d^2E}{d\omega dt}(\omega)$: varying $Q\in\{0.1,\ldots,0.6\}$; 
    $\alpha=0.01,\,\kappa=1,\,M=1$
\end{minipage}

\vspace{-0.2cm}

\caption{Behavior of the energy emission rate for the tree spins. The angular quantum number $\kappa=1$ is used throughout.}
\label{Ec}
\end{figure*}

Moreover, we should note that the three spin sectors have the same general shape \ref{Ec}, where the emission starts from zero at very small frequency, rises to a maximum, and then decays as the thermal factor suppresses high-frequency quanta.  This behavior is a direct consequence of the product between the greybody lower bound and the Hawking distribution.  At low \(\omega\), the barrier factor is dominant, blocking the emission.  At high \(\omega\), the exponential thermal denominator becomes dominant.  The maximum occurs in the intermediate region where the two effects balance.

The relative magnitudes of the spectra are governed by both statistics and barrier penetration.  The scalar case is the most geometrically sensitive because \(V_0\) contains \(f'(r)/r\).  This extra curvature term increases the effective barrier for the parameters used in the notebook, making the scalar emission lower than the spin-one emission.  The electromagnetic channel depends primarily on the centrifugal barrier, so it is less suppressed by the metric derivative. The spinor channel, in the approximation implemented here, has the smallest barrier integral among the three cases for the chosen angular numbers. This explains why the spin-\(1/2\) curves can appear with larger amplitude even though the Fermi-Dirac distribution does not contain the low-frequency Bose enhancement.

Therefore, based on the analysis in this section, we arrive at a useful physical reading. In this context, the spin of the emitted field determines how strongly the field couples to the effective potential outside the horizon. The black hole parameters determine the horizon radius and the temperature. The observed spectrum results from both effects simultaneously. A hotter black hole does not necessarily emit much more in a given channel if the greybody barrier also grows, while a lower barrier may still produce a suppressed flux if the Hawking temperature is too small. The plots show that, for the chosen parameter sets, the temperature effect is decisive when varying \(\alpha\), whereas the cooling effect is decisive when varying \(Q\).

\section{Conclusions}\label{s5}

In this work, we investigated the propagation of massless fields with spins $s=0$, $s=1$, and $s=1/2$ in the background of an EH black hole surrounded by PFDM. The geometry considered here combines two physically motivated corrections to the standard charged black hole scenario. The first is related to the nonlinear electrodynamic contribution in the EH effective theory, while the second is due to the logarithmic correction induced by the surrounding perfect-fluid dark-matter distribution. This framework provides a useful setting for analyzing how quantum-electrodynamic nonlinearities and dark matter effects modify the scattering properties of black holes.

We first reviewed the main properties of the metric function and showed how the EH parameter, the electric charge, and the PFDM parameter affect the horizon structure. The presence of the PFDM contribution changes the radial profile of the metric function through a logarithmic term, while the EH correction introduces higher-order inverse powers of the radial coordinate. As a consequence, the effective geometry differs from both the Schwarzschild and Reissner-Nordstr\"om limits, particularly in the region near the horizon, where the scattering potential is most sensitive to the background parameters.

For each spin sector, we derived the corresponding Schr\"odinger-like radial equation and constructed the effective potential governing wave propagation. The scalar, electromagnetic, and Dirac fields probe the geometry in different ways, leading to distinct potential barriers and, consequently, different transmission probabilities. In all cases, the greybody factors were estimated using the Boonserm-Visser bound method, which provides a rigorous analytical lower bound for the transmission coefficient expressed as integrals over the effective potential. This approach allowed us to study the dependence of the transmission spectra and absorption cross sections on the black hole parameters without requiring an exact analytical solution of the scattering equation.

Our results show that the greybody lower bounds increase monotonically with the frequency $\omega$, approaching unity in the high-frequency regime, as expected for waves with enough energy to overcome the effective potential barrier. At low frequencies, however, the transmission probability is strongly suppressed, and the details of the geometry become more relevant. The electric charge, the EH parameter, and the PFDM parameter modify the height and width of the effective potential barrier, thereby changing the probability that radiation escapes to infinity. In particular, variations in these parameters can either enhance or suppress transmission, depending on how they shift the event horizon and deform the effective potential.

The comparison among the different spin sectors reveals a clear spin-dependent behavior. For the same background parameters and angular quantum numbers, the Dirac field generally exhibits a different transmission profile from the scalar and electromagnetic fields. This occurs because the spin-$1/2$ potential contains terms associated with the Dirac angular quantum number $\kappa$, while the scalar and electromagnetic potentials depend on the orbital number $l$ in different ways. Consequently, the absorption cross sections also display distinct magnitudes and frequency dependence for different spins. These results indicate that greybody lower bounds are sensitive probes not only of the background geometry but also of the intrinsic spin of the perturbing field.

We also compared limiting cases of the model, including the Schwarzschild, Reissner-Nordström, Euler--Heisenberg, and EH plus PFDM geometries. This comparison makes explicit how each physical ingredient contributes to the final scattering behavior. The Schwarzschild case provides the reference profile; the charged Reissner-Nordström geometry introduces electromagnetic corrections; the EH contribution encodes nonlinear electrodynamic effects; and the PFDM term accounts for the influence of the surrounding dark matter distribution. The combined EH+PFDM case, therefore, contains a richer parameter dependence and leads to potentially distinguishable changes within the model in the greybody bounds and absorption spectra.

From a physical perspective, our analysis shows that the black-hole greybody lower bounds can serve as useful diagnostics for distinguishing between different modifications of the standard charged black-hole geometry. Since these quantities directly affect the spectrum of Hawking radiation observed at infinity, any deformation of the effective potential caused by nonlinear electrodynamics or dark matter may leave an imprint on the evaporation process and on the associated absorption cross section. Although the present study is based on analytical bounds rather than exact numerical transmission coefficients, the results provide a consistent and transparent picture of how the EH and PFDM parameters influence wave propagation.

There are several possible extensions of this work. One natural continuation is to compute the exact greybody factors by direct numerical integration of the radial wave equations and compare the exact transmission coefficients with the Boonserm-Visser lower bounds obtained here. Another interesting direction is to analyze the corresponding quasinormal mode spectrum and investigate its connection with the structure of the effective potential. It would also be worth extending the present study to massive fields, higher angular modes, rotating backgrounds, and alternative dark matter profiles. Such investigations may help clarify the extent to which greybody factors, absorption cross sections, shadows, and quasinormal modes can jointly constrain, or at least phenomenologically probe, nonlinear electrodynamic corrections and dark matter effects in black-hole spacetimes.

Therefore, the black hole geometry discussed in this work provides a rich and analytically tractable background for studying black hole scattering. The results obtained in this paper demonstrate that both the nonlinear electromagnetic sector and the surrounding dark matter distribution have nontrivial effects on the effective potentials, greybody bounds, and absorption cross sections of massless fields. These findings reinforce the role of greybody bounds as sensitive probes of modified black hole geometries and motivate further studies of wave propagation in black hole backgrounds beyond the standard Einstein-Maxwell framework.

\section*{Acknowledgments}

\hspace{0.5cm} The author Fernando Belchior would like to to express gratitude to the Conselho Nacional de Desenvolvimento Cient\'{i}fico e Tecnol\'{o}gico CNPq for grant No. 151845/2025-5. The author Edilberto Silva acknowledges the support from Conselho Nacional de Desenvolvimento Cient\'{i}fico e Tecnol\'{o}gico (CNPq) (grants 306308/2022-3), Funda\c{c}\~{a}o de Amparo \`{a} Pesquisa e ao Desenvolvimento Cient\'{i}fico e Tecnol\'{o}gico do Maranh\~{a}o (FAPEMA) (grants UNIVERSAL-06395/22), and Coordena\c{c}\~{a}o de Aperfei\c{c}oamento de Pessoal de N\'{i}vel Superior (CAPES) - Brazil (Code 001). The author Faizuddin Ahmed acknowledges the Inter University Centre for Astronomy and Astrophysics (IUCAA), Pune, India for granting visiting associateship.

%


\end{document}